# Uncertainty-Aware Cross-Modal Remote Sensing Image–Text Retrieval via Evidential Learning

Zhuoyue Wang, Xueqian Wang, *Member, IEEE*, Gang Li, *Senior Member, IEEE*,
Chengxi Li, *Member, IEEE*, Yongpan Liu, *Senior Member*, *IEEE*, and Yifang Ban, *Senior Member, IEEE*

***Abstract*—In cross-modal remote sensing image–text retrieval (CMRSITR), test-time remote sensing (RS) images and textual descriptions may deviate from well-curated benchmark conditions because of sensor- and atmosphere-related image degradations and text-side RS-vocabulary heterogeneity. Under such non-ideal conditions, existing CMRSITR methods may produce unreliable retrieval results because they usually perform retrieval with full certainty for each query and do not distinguish the varying uncertainty introduced across queries. To address this issue, we propose an evidential learning-based CMRSITR (ELC) method for uncertainty-aware retrieval. During the training phase of ELC, evidential learning (EDL) is employed to model the inter-modal correspondences between RS images and textual descriptions as Dirichlet distributions, from which the uncertainty of each query can be obtained. Based on the EDL outputs, uncertainty–correctness alignment learning (UCL) is introduced to align the estimated uncertainty with retrieval correctness, encouraging high uncertainty for incorrect retrieval and low uncertainty for correct retrieval. Furthermore, intra-modal relationship learning (RL) distills the intra-modal similarity structure from pretrained mentor encoders for the trainable encoders, thereby making the Dirichlet distributions modeled by EDL more discriminative. In the test phase of ELC, the estimated uncertainty is compared with a threshold determined by a fixed deferral ratio, where low-uncertainty queries are directly returned and high-uncertainty queries are refined by RS-aware test-time augmentation (RS-TTA). Experimental results demonstrate that our proposed ELC achieves competitive retrieval performance in comparison with existing state-of-the-art CMRSITR methods and provides stronger robustness under the evaluated RS-specific degradations, including sensor- and atmosphere-related image perturbations and RS-vocabulary heterogeneity.**

This work was supported by National Key R&D Program of China under Grant 2025YFF0514602, National Natural Science Foundation of China under Grants 62471274, 62341130, and Beijing Natural Science Foundation under Grant JQ2501. Part of this paper has been published by International Geoscience and Remote Sensing Symposium 2025 [1]. Corresponding Author: Xueqian Wang.

Z. Wang and Y. Liu are with the Department of Electronic Engineering, Tsinghua University, Beijing 100084, China.

X. Wang and G. Li are with the Department of Electronic Engineering, Tsinghua University, Beijing 100084, China, and also with the State Key Laboratory of Space Network and Communications, Tsinghua University, Beijing 100084, China (e-mail: wangxueqian@mail.tsinghua.edu.cn).

C. Li is with the School of Electrical Engineering and Computer Science, KTH Royal Institute of Technology.

Y. Ban is with the Division of Geoinformatics, School of Architecture and the Built Environment, KTH Royal Institute of Technology.

## I. INTRODUCTION

With the rapid development of Earth observation technology [2]-[4], an enormous volume of remote sensing (RS) data, including images and textual descriptions, is continuously generated by numerous satellite sensors. RS images provide rich spatial and spectral details but often lack explicit semantic meaning. In contrast, textual descriptions provide semantic context and high-level interpretations but may lack detailed spatial precision. Effectively leveraging both visual and textual modalities is crucial for improving the accessibility and interpretability of RS data, which is essential for facilitating large-scale geospatial analysis such as environmental monitoring [2] and disaster response [5]. To this end, a fundamental task is cross-modal RS image–text retrieval (CMRSITR). CMRSITR aims to bridge RS images and textual descriptions, enabling the retrieval of relevant RS images based on textual queries and vice versa.

Recent progress in deep learning has notably improved CMRSITR performance [4]. Typically, deep neural network (DNN)-based methods are designed to learn a shared embedding space in which image and text features can be projected and directly compared, enabling effective and scalable cross-modal similarity computation. The existing DNN-based methods can be broadly categorized into feature extraction structure design [6]-[21] and embedding learning paradigms [22]-[36]. DNN-based methods emphasizing feature extraction aim to generate discriminative representations for CMRSITR by employing techniques such as multi-scale feature fusion [6]-[8] and attention mechanisms [10]-[14], commonly realized through the novel design of DNN structures. Embedding learning methods concentrate on developing novel learning paradigms to produce better aligned embeddings with enhanced retrieval performance, e.g., contrastive learning techniques [22]-[24] and the adaptation of pretrained vision-language models to RS tasks [25], [26]. These existing approaches have achieved notable performance on benchmark datasets and form the basis of most current CMRSITR systems.

Despite the notable progress achieved by current CMRSITR methods, most of them are designed for ideal test datasets, where both images and textual descriptions are carefully preprocessed to remove possible noise. In practical applications, however, it is difficult for the test RS data to fulfill these idealized conditions. For optical RS images, a major source of uncertainty arises from atmospheric and acquisition conditions, including thin-cloud or haze attenuation, as well as illumination variation and radiometric drift across acquisitions [37]-[39]. Despite substantial improvements in modern optical satellite imaging quality, residual sensor-physics artifacts, such as readout Gaussian fluctuations and row-wise push-broom striping, may persist in real acquisitions and degrade the reliability of fine-grained visual details in applications [40]-[43]. In parallel, text queries in CMRSITR may suffer from RS-vocabulary heterogeneity across captioning campaigns. For example, terminology may drift among semantically related RS terms such as 'harbor', 'dock', and 'marina' while captions may also differ in granularity when describing land-cover patterns, object density, or spatial layouts [44], [45]. In such noisy conditions, retrieval performance of existing CMRSITR models often degrades considerably [45]. This is because existing methods implicitly assume that the trained model could perform retrieval with full certainty for each query and they neglect the fact that noise in test data introduces varying levels of uncertainty across queries.

In this paper, we propose a new **e**vidential **l**earning-based **C**MRSITR (ELC) method. During training of ELC, our method incorporates evidential learning (EDL) to model RS image–text correspondence and estimate uncertainty for each query, while an uncertainty-aware weighting (UW) strategy down-weights high-uncertainty samples in the EDL loss to suppress ambiguous correspondences. Based on the outputs of EDL, we introduce an uncertainty–correctness alignment learning (UCL) loss, which encourages high uncertainty to be associated with low retrieval accuracy. Through this design, the model learns uncertainty estimates aligned with retrieval correctness, providing a reliability signal for selective retrieval, thereby extending the conventional CMRSITR pipeline with a principled uncertainty-aware inference mechanism. Furthermore, intra-modal relationship learning (RL) leverages pretrained mentor encoders to distill intra-modal similarity structure into the trainable encoders, producing more discriminative features and thereby supporting more informative Dirichlet outputs at the EDL head. During the test stage, ELC compares the estimated uncertainty with a threshold determined by a fixed deferral ratio, and then low-uncertainty queries are returned directly, while high-uncertainty queries are deferred to an RS-aware test-time augmentation (RS-TTA) stage. By aggregating retrieval results from multiple augmented versions of the query, RS-TTA helps suppress the influence of noise and stabilizes the predictions in uncertain conditions. Unlike standard CMRSITR approaches that output only ranked retrievals, ELC supports selective inference by refining high-uncertainty queries (while directly returning low-uncertainty queries).

The main contributions are summarized as follows:

1) Our ELC for CMRSITR incorporates UCL and RL during training, where UCL improves the uncertainty-awareness estimated by EDL and RL enhances the learning capability of EDL by capturing intra-modal ranking relationships.
2) During the test phase, ELC utilizes the estimated uncertainty to assess the reliability of each query. For highly uncertain inputs, we design RS-TTA to refine the retrieval result through aggregated evaluation. This adaptive mechanism allows the system to stabilize predictions and suppress the impact of noisy inputs, providing stronger robustness under noisy conditions.
3) Experimental results underscore that our ELC shows more pronounced gains under RS-specific test-time degradations, including sensor- and atmosphere-related RS image perturbations and text-side RS-vocabulary heterogeneity in comparison with existing state-of-the-art methods.

The rest of this paper is organized as follows. Section II presents a review of related work on CMRSITR, EDL, and RS-TTA. Section III provides a detailed description of the proposed ELC method. Section IV provides in-depth experimental comparisons between the proposed ELC and existing methods. Finally, Section V concludes this article with several remarks on possible future research directions.

## II. Related Work

### *A. Cross-Modal Remote Sensing Image–Text Retrieval*

CMRSITR aims to retrieve the most relevant image for a given text description, or vice versa, from a large-scale RS dataset. Existing CMRSITR methods can be broadly categorized into two groups: feature extraction structure design [6]-[21], which focuses on building effective architectures to capture rich representations, and embedding learning paradigms [22]-[36], which focus on designing effective feature alignment strategies after feature extraction.

Feature extraction structure design methods aim to generate discriminative cross-modal representations by designing novel DNN architectures [6]-[21]. Multi-scale and local-global fusion approaches [6]-[9], [16], [21] address the wide scale variation of RS objects: Yuan *et al.* [6] propose a multi-scale fine-grained retrieval algorithm targeting the small-object regime of RS images, and Hu *et al.* [8] couple global and local features across modalities through soft alignment. Several methods employ transformer architectures to model cross-modal relationships by leveraging attention mechanisms [10]-[14], as in the interaction-enhancing transformer of Tang *et al.* [10] and the spatial-channel attention transformer of [12]. In addition, graph-based learning models have also been introduced into the feature extraction stage to capture structured intra-modal relationships and enhance cross-modal feature alignment [46]-[48].

Different from methods that focus on feature extraction structure design, several methods aim to enhance cross-modal semantic alignment by developing effective embedding learning paradigms [22]-[36]. A prominent direction is contrastive learning, which leverages discriminative objectives to narrow the modality gap. Zhao *et al.* [23] integrate dynamic contrastive learning with a masked vision-language modeling strategy to enhance token-level alignment and global image–text representation consistency. Curriculum-based contrastive learning has been adopted to progressively refine representation alignment on hyperspherical manifolds [24]. Beyond contrastive paradigms, some methods augment matching loss with auxiliary supervision to inject structure that captions alone do not carry [18], [29], [30], [33]-[35]. Mi *et al.* [29] introduce knowledge graphs to enhance textual semantics and bridge cross-modal gaps. More recently, pretrained vision-language models [31] have been introduced into CMRSITR, where large-scale models trained on natural image–text pairs are adapted to the RS domain. These methods typically retain the pretrained backbone and finetune it on domain-specific RS data to enhance modality alignment and improve retrieval performance [25], [26]. RemoteCLIP [25] illustrates this route by continually pretraining CLIP on a unified large-scale RS image-text dataset and reporting substantial retrieval gains, while parameter-efficient adaptation-prompt tuning [27], [28], [31], and adapter-based methods [32], [36] have recently emerged as efficient approaches for adapting pretrained models without modifying their original parameters.

Although the aforementioned CMRSITR methods have achieved impressive results on benchmark datasets, they inherently assume well-curated test conditions. In practical applications, queries inevitably suffer from RS-specific test-time degradations stemming from sensor- and atmosphere-physics and caption-side annotation heterogeneity. Most existing methods overlook the pervasive RS-specific degradations under non-ideal noisy conditions, rendering their confident predictions unreliable and causing considerable performance degradation.

### *B. Evidential Learning*

Uncertainty estimation in DNN-based models quantifies the reliability and confidence of model predictions. The Bayesian neural network [49] is a classical approach for uncertainty estimation, which models uncertainty through probabilistic distributions over model weights. Monte Carlo dropout [50] approximates Bayesian inference through multiple stochastic forward passes, and deep ensembles [51] aggregate independently trained models. These approaches share a common limitation: high computational cost and sensitivity to weight-prior selection [49]-[51]. Recently, EDL has been introduced as an efficient uncertainty quantification framework derived from Dempster-Shafer Theory [52] and Subjective Logic [53]. Neural networks directly output evidence parameterizing Dirichlet distributions over class probabilities within EDL, enabling uncertainty estimation in a single forward pass [54]. EDL has been widely applied across various tasks beyond its original formulation for classification [54], [55], including domain adaptation [56], temporal action localization [57], and cross-modal retrieval [58]. Although EDL provides an efficient way to estimate uncertainty in a single forward pass, directly applying vanilla EDL to CMRSITR remains suboptimal. Vanilla EDL estimates uncertainty mainly from the total amount of Dirichlet evidence, but this uncertainty is not necessarily aligned with retrieval correctness and therefore cannot directly serve as a retrieval-correctness reliability signal for selective retrieval. Moreover, vanilla EDL models inter-modal correspondence but does not explicitly preserve intra-modal relationships among RS images or captions. In CMRSITR, these relationships are informative because visually similar scenes often share land-cover patterns, object density, and spatial layouts, while semantically related captions may describe the same scene at different granularities. Ignoring such structure makes the learned image and text features less discriminative, which weakens the similarity evidence used by the EDL head. To address these limitations, our ELC introduces UCL to align estimated uncertainty with retrieval correctness and RL to improve feature discriminability for more reliable EDL-based uncertainty learning.

### *C. Test-Time Augmentation*

Test-time augmentation (TTA) is an inference-stage technique that improves predictive accuracy and robustness by aggregating predictions from transformed versions of a test sample [59], [60]. It has been widely used in computer vision tasks such as image classification [61], semantic segmentation [62], as well as in natural language processing for tasks such as text classification [63]. However, directly applying generic TTA to CMRSITR is insufficient because the validity of augmentation operators depends on the radiometric, geometric, and semantic characteristics of RS data. For optical RS images, unconstrained color jitter does not follow the radiometric-calibration form of satellite acquisitions, and horizontal flipping may alter orientation-dependent spatial relations described by RS captions. For textual queries, generic synonym replacement such as easy data augmentation (EDA) [64] may introduce out-of-domain substitutions that do not preserve RS-specific semantics. Moreover, generic TTA is usually applied to all test queries, whereas noisy CMRSITR requires query-level treatment because uncertainty varies across queries. These observations motivate an RS-TTA strategy in ELC, where radiometric, small-angle rotation, and RS-vocabulary augmentations are selectively applied according to estimated uncertainty.

This paper significantly extends the content of its conference version [1] by linking estimated per-query uncertainty to test-time retrieval, so that uncertainty is used not only to characterize retrieval reliability but also to identify queries requiring refined analysis under non-ideal noisy conditions. Experiments regarding RS-specific test-time degradations arising from sensor- and atmosphere- effects and caption-side heterogeneity are also substantially expanded to show the effectiveness of our ELC.

## III. Proposed Method

In this section, we introduce the implementation of the training and test stages of the proposed ELC method. CMRSITR queries are often affected by two RS-specific degradation sources: sensor- and atmosphere-physics perturbations and caption-side annotation heterogeneity. These challenges motivate ELC, a selective retrieval framework with uncertainty-aware capability. Specifically, ELC first employs EDL to produce a per-query Dirichlet uncertainty estimate (Section III-B), then incorporates UCL to align the uncertainty estimate with retrieval correctness (Section III-C). RL further serves as an auxiliary component that distills mentor-side intra-modal relational structure into the EDL branch (Section III-D). Based on the estimated uncertainty, RS-TTA selectively defers only high-uncertainty queries for radiometric, small-angle rotation, and RS-vocabulary augmentations (Section III-E).

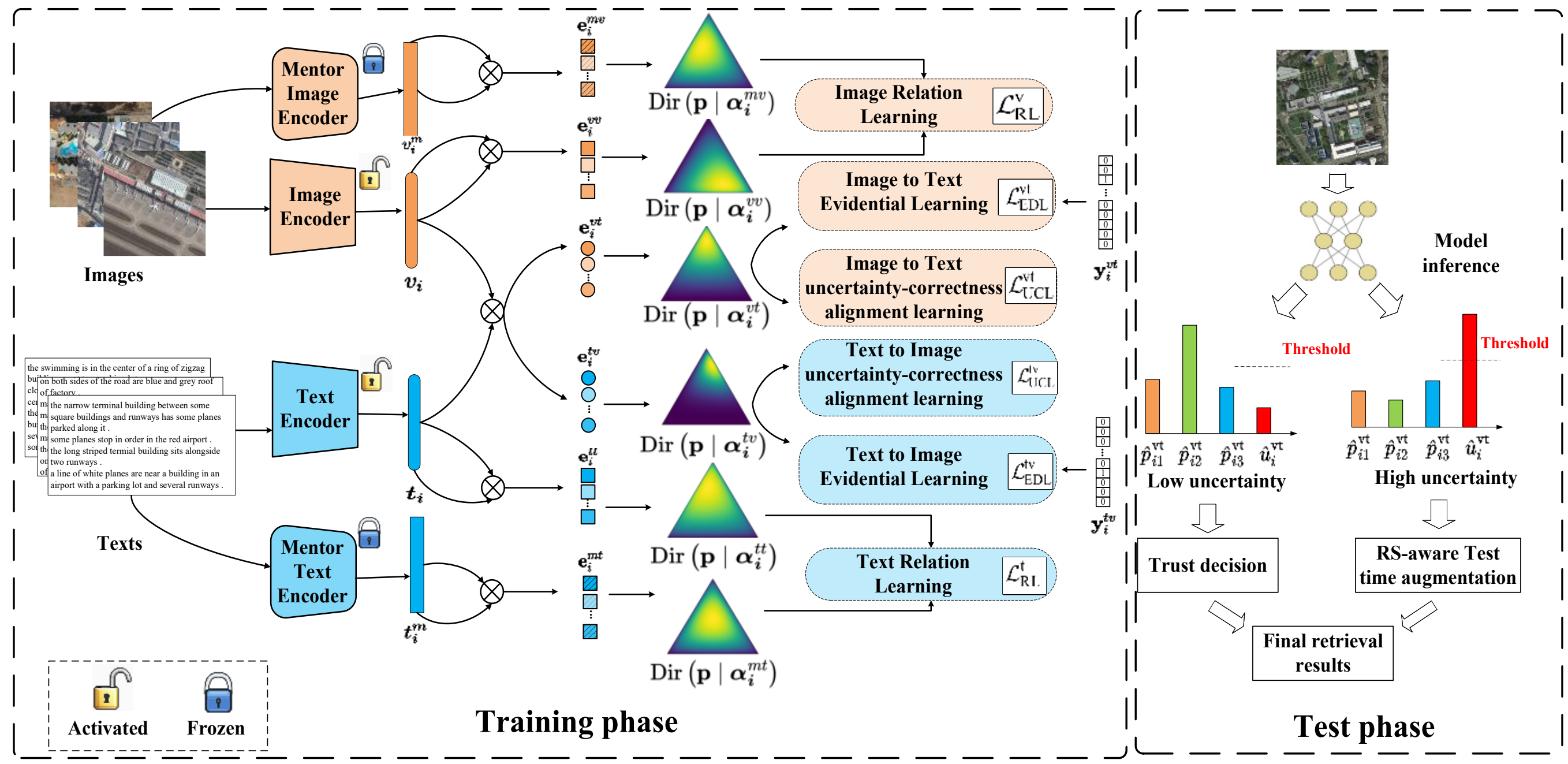

Fig. 1. Overall architecture of the proposed ELC method. In the training stage, dual-stream encoders (including trainable encoders and frozen mentor encoders) extract features from image-text pairs. Evidential representations are constructed and modeled as Dirichlet distributions to represent uncertainty, and the network is jointly optimized through EDL, UCL, and RL. In the test stage, a query sample is evaluated based on its predictive uncertainty. Queries with low uncertainty are directly trusted for retrieval, whereas high-uncertainty samples are further refined through RS-TTA to improve retrieval reliability.

The RS image–text pair dataset is denoted by $\mathcal{D}=\{(V_i,T_i)\,|\,i=1,2,...,n\}$, where $V_i$ and $T_i$ represent the $i$-th RS image and its corresponding textual description, respectively, and $n$ denotes the size of the dataset. The RS image and text datasets are denoted by $\mathcal{V}=\{v_i\,|\,i=1,2,...,n\}$ and $\mathcal{T}=\{T_i\,|\,i=1,2,...,n\}$, respectively. The CMRSITR task is to retrieve semantically relevant samples in one modality (e.g., texts) from a large-scale RS dataset, given queries in the other modality (e.g., RS images). The diagram of the training and test stages of our proposed method are shown in Fig. 1.

*A. Overview of the Training Stage*

The training stage of our ELC jointly optimizes EDL, UCL, and RL over batched RS image–text pairs as detailed below. During the training stage, we describe the training process using a batch of $K$ training data samples, where each sample consists of a pair of image $V_i$ and text $T_i$, for $i=1,2,...,K$. First, all the image and text samples are fed into the image encoder and the text encoder, respectively, producing the corresponding image features $\mathbf{v}_i$ and text features $\mathbf{t}_i$. Next, the similarity scores between each image feature and text feature are computed, yielding $\{\langle\mathbf{v}_i,\mathbf{t}_j\rangle|\ i,j=1,\ldots,K\}$, where $\langle\cdot\rangle$ denotes the inner product operation. Meanwhile, the self-similarity within each modality is evaluated based on the similarity scores between pairs of image features and pairs of text features, resulting in $\{\langle\mathbf{v}_i,\mathbf{v}_j\rangle|\ i,j=1,\ldots,K\}$. Based on these similarity measures, EDL, UCL, and RL are employed to model the inter-modal correspondences and the intra-modal relationships.

*B. Inter-Modal Evidential Learning*

EDL involves both image-to-text modeling and text-to-image modeling of the inter-modal correspondences. For the former, given any image feature $\mathbf{v}_i$, we define a vector consisting of the similarity scores between this image feature and all text features in the batch as follows:

$$\mathbf{s}_i^{\text{vt}}=\left[\langle\mathbf{v}_i,\mathbf{t}_1\rangle,\langle\mathbf{v}_i,\mathbf{t}_2\rangle,\ldots,\langle\mathbf{v}_i,\mathbf{t}_K\rangle\right]^{\text{T}},\forall i. \tag{1}$$

Based on Eq. (1), the similarity scores $\mathbf{s}_i^{\text{vt}}$ are input into the EDL activation function $\text{EDL}_{act}(\cdot)=\exp(\cdot)$ to generate

$$\mathbf{e}_i^{\text{vt}}=\text{EDL}_{\text{act}}\left(\mathbf{s}_i^{\text{vt}}\right), \tag{2}$$

where $\mathbf{e}_i^{\text{vt}}$ is a column vector of length $K$. Here, $\text{EDL}_{\text{act}}(\cdot)=\exp(\cdot)$ denotes the exponential activation, ensuring the output evidence vector is non-negative. Then, the distribution of the image-to-text correspondence for image $V_i$ is denoted by [54],

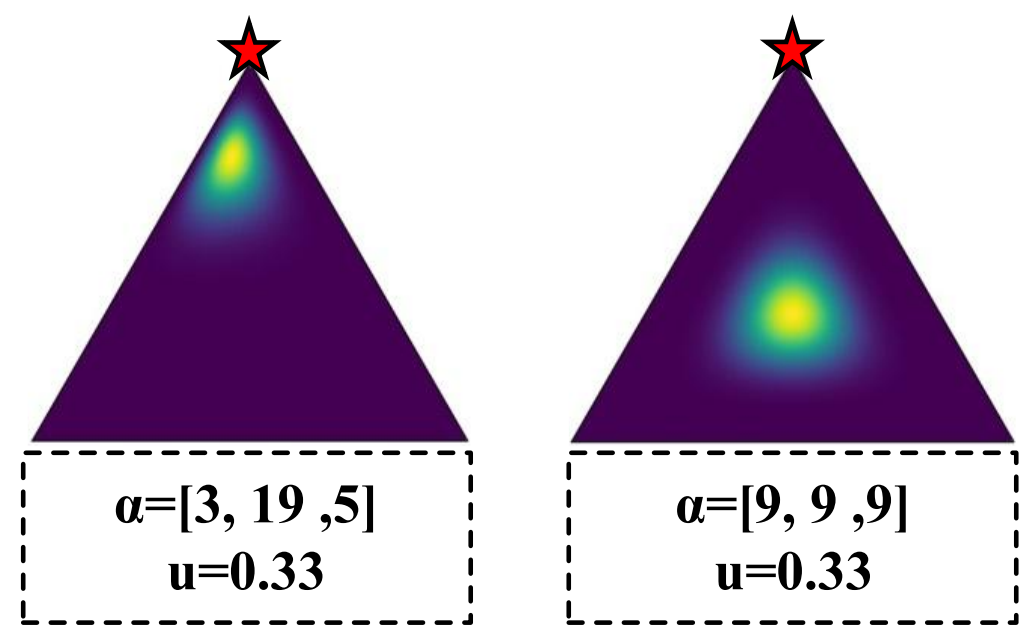

Fig. 2. Illustration of two Dirichlet distributions on the 3-class simplex sharing the same uncertainty but qualitatively different. Each vertex of the triangle denotes a pure class, and each point inside denotes a probability vector. Heatmap intensity encodes the Dirichlet probability density (lighter = higher). The pentagram marks the ground-truth class. Despite identical uncertainty, the left distribution concentrates near the correct class (yielding a correct decision), while the right distribution is misaligned (yielding an ambiguous decision). This motivates UCL, which encourages high uncertainty for incorrect retrieval and low uncertainty for correct retrieval.

$$\mathrm{Dir}\left(\mathbf{p} \mid \boldsymbol{\alpha}_i^{\mathrm{vt}}\right)=\begin{cases}\dfrac{1}{\mathrm{B}\left(\boldsymbol{\alpha}_i^{\mathrm{vt}}\right)}\displaystyle\prod_{j=1}^{K} p_j^{\alpha_{ij}^{\mathrm{vt}}-1}, & \text{for } \mathbf{p} \in \mathcal{SP}_K \\ 0, & \text{otherwise}\end{cases}, \quad (3)$$

where $\mathrm{B}(\cdot)$ represents the $K$-dimensional beta function, $\left\{\boldsymbol{\alpha}_i^{\mathrm{vt}}=\mathbf{e}_i^{\mathrm{vt}}+\mathbf{1}, i=1,\ldots,K\right\}$ denotes the concentration parameters with $\boldsymbol{\alpha}_i^{\mathrm{vt}}=\left[\alpha_{i1}^{\mathrm{vt}},\ldots,\alpha_{iK}^{\mathrm{vt}}\right]$, and $\mathbf{1}$ is an all-one column vector. In Eq. (3), $\mathcal{SP}_K$ is the $K$-dimensional unit simplex defined as

$$\mathcal{SP}_K=\left\{\mathbf{p} \,\Big|\, \sum_{i=1}^{K} p_i=1 \text{ and } 0 \le p_1,\ldots,p_K \le 1\right\}. \quad (4)$$

In Eqs. (3) and (4), a larger value of $p_j$ (where $p_j$ is a latent simplex variable and all loss terms below use closed-form Dirichlet expectations and do not require sampling $p_j$ ) indicates that image $V_i$ and text $T_j$ are of higher image-to-text correspondence, $\forall j$ . Note that the true label of image-to-text correspondence can be expressed as $\mathbf{y}_i^{\mathrm{vt}}=[y_{i1}^{\mathrm{vt}}, y_{i2}^{\mathrm{vt}},\ldots,y_{iK}^{\mathrm{vt}}]^{\mathrm{T}}$, where $y_{ij}^{\mathrm{vt}}=1$ for $i$=$j$ and $y_{ij}^{\mathrm{vt}}=0$ otherwise.

According to the property of the Dirichlet distribution in Eq. (3), the uncertainty of the image-to-text correspondence can be measured by $u_i^{\mathrm{vt}}=K/S_i^{\mathrm{vt}}$ , where $S_i^{\mathrm{vt}} \triangleq \sum_{j=1}^{K}\alpha_{ij}^{\mathrm{vt}}$ and a lower value of $u_i^{\mathrm{vt}}$ indicates a higher level of confidence. During training, samples with high predictive uncertainties have ambiguous image-text correspondence. Treating such ambiguous samples with equal importance as others may mislead the training process. Directly applying the negative log-likelihood over all in-batch samples treats ambiguous pairs equally with reliable ones, which can bias the gradients toward noisy correspondences. To mitigate this issue, we introduce a UW strategy into the EDL loss, where high-uncertainty samples are assigned smaller weights. This design reduces the adverse impact of ambiguous correspondences, allowing the model to focus on more reliable pairs and enhancing retrieval robustness.

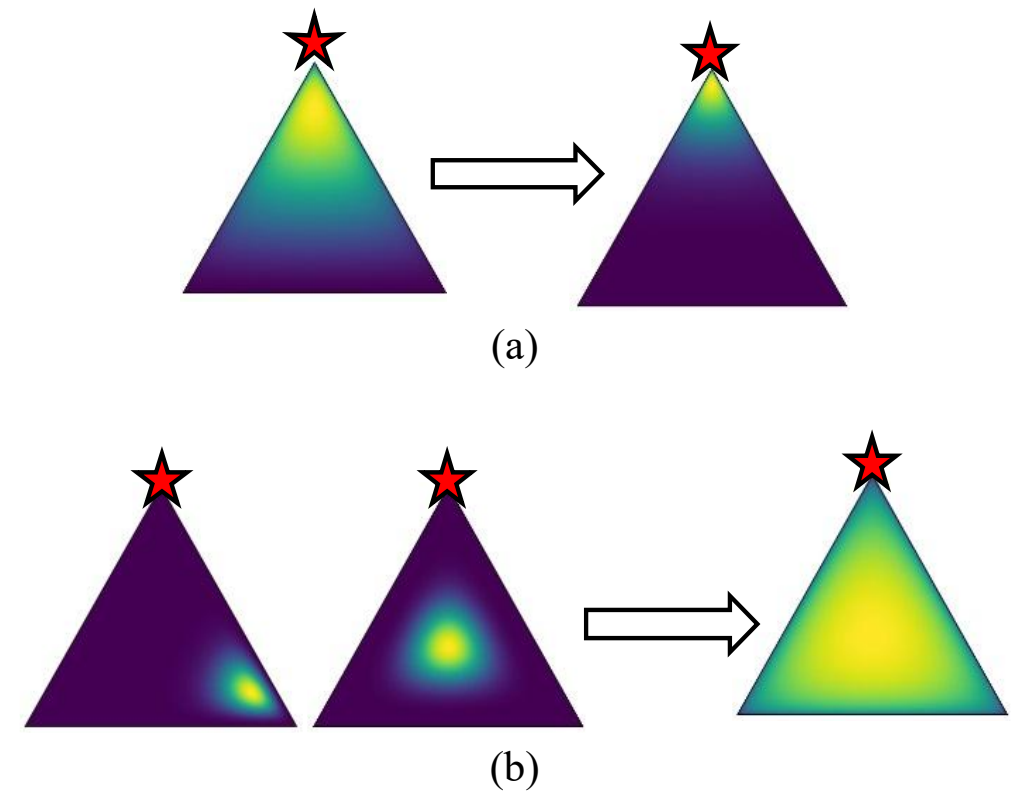

Fig. 3. Illustration of uncertainty–correctness alignment in UCL. The pentagram denotes the correct decision point. (a) For correct decisions, UCL encourages lower uncertainty. (b) For incorrect decisions, UCL encourages higher uncertainty.

Incorporating the proposed uncertainty-aware weighting strategy, the negative log-likelihood loss for the image-to-text correspondence over all the images in the batch is given as

$$\begin{aligned}\mathcal{L}_{\mathrm{EL}}^{\mathrm{vt}} &= \frac{1}{K}\sum_{i=1}^{K} -(1-u_i^{\mathrm{vt}})\log\left(\mathrm{L}(\mathbf{y}_i^{\mathrm{vt}} \mid \boldsymbol{\alpha}_i^{\mathrm{vt}})\right) \\ &= \frac{1}{K}\sum_{i=1}^{K}(1-u_i^{\mathrm{vt}})\sum_{j=1}^{K} y_{ij}^{\mathrm{vt}}\left(\log\left(S_i^{\mathrm{vt}}\right)-\log\left(\alpha_{ij}^{\mathrm{vt}}\right)\right),\end{aligned} \quad (5)$$

where the likelihood function for image $V_i$ is expressed as

$$\mathrm{L}(\mathbf{y}_i^{\mathrm{vt}} \mid \boldsymbol{\alpha}_i^{\mathrm{vt}})=\int_{\mathbf{p}\in\mathcal{SP}_K}\prod_{j=1}^{K} p_j^{y_{ij}}\frac{1}{B\left(\boldsymbol{\alpha}_{\mathrm{i}}^{\mathrm{vt}}\right)}\prod_{j=1}^{K} p_j^{\alpha_{ij}^{\mathrm{vt}}-1} d\mathbf{p}, \quad (6)$$

For unpaired image-text samples, we aim to make the distribution of image-to-text correspondence close to a uniform distribution, reflecting zero evidence for any specific pairing. To enforce this uniformity constraint, the Kullback–Leibler (KL) divergence term is incorporated into the training loss as follows:

$$\mathcal{L}_{\mathrm{KL}}^{\mathrm{vt}}=\frac{1}{K}\sum_{i=1}^{K}\mathrm{KL}\left[\mathrm{Dir}\left(\mathbf{p}_i \mid \tilde{\boldsymbol{\alpha}}_i^{\mathrm{vt}}\right) \| \mathrm{Dir}\left(\mathbf{p}_i \mid \mathbf{1}\right)\right], \quad (7)$$

where $\tilde{\boldsymbol{\alpha}}_i^{\mathrm{vt}}=\mathbf{y}_i^{\mathrm{vt}}+\left(1-\mathbf{y}_i^{\mathrm{vt}}\right)\odot\boldsymbol{\alpha}_i^{\mathrm{vt}}$ denotes the concentration parameters with misleading evidence, $\odot$ denotes element-wise multiplication, and $\mathrm{KL}\left(\cdot\|\cdot\right)$ denotes the KL divergence between two probabilistic distributions. By minimizing the loss in Eq. (7), the concentration parameters $\tilde{\boldsymbol{\alpha}}_i^{\mathrm{vt}}$ with misleading evidence produced by the EDL activation function are encouraged to approach the all-one vector (i.e., the uniform Dirichlet prior), so that misleading evidence does not contribute to the predicted correspondence. During training, this mechanism mitigates the misleading effects of samples lacking image-text correspondences. In cases where the in-batch text features cannot reliably support any image-to-text correspondence for image $V_i$, the EDL head is encouraged to produce uninformative (near-uniform Dirichlet) outputs.

Finally, based on Eqs. (5) and (7), the EDL training loss for image-to-text modeling can be expressed as

$$\mathcal{L}_{\mathrm{EDL}}^{\mathrm{vt}} = \mathcal{L}_{\mathrm{EL}}^{\mathrm{vt}} + \frac{n_{\mathrm{e}}}{b_1}\mathcal{L}_{\mathrm{KL}}^{\mathrm{vt}}, \tag{8}$$

where $b_1$ is a hyperparameter that controls the regularization effect of the KL divergence, and $n_{\mathrm{e}}$ denotes the index of the current epoch. The KL regularization is annealed by the epoch-dependent factor $n_{\mathrm{e}} / b_1$, following standard EDL practice [54]. Likewise, the EDL training loss for text-to-image modeling is given as

$$\mathcal{L}_{\mathrm{EDL}}^{\mathrm{tv}} = \mathcal{L}_{\mathrm{EL}}^{\mathrm{tv}} + \frac{n_{\mathrm{e}}}{b_1}\mathcal{L}_{\mathrm{KL}}^{\mathrm{tv}}, \tag{9}$$

where the calculation of $\mathcal{L}_{\mathrm{EL}}^{\mathrm{tv}}$ and $\mathcal{L}_{\mathrm{KL}}^{\mathrm{tv}}$ follows a procedure similar to that of $\mathcal{L}_{\mathrm{EL}}^{\mathrm{vt}}$, and $\mathcal{L}_{\mathrm{KL}}^{\mathrm{vt}}$. According to Eqs. (8) and (9), the overall EDL training loss can be expressed as

$$\mathcal{L}_{\mathrm{EDL}} = \mathcal{L}_{\mathrm{EDL}}^{\mathrm{vt}} + \mathcal{L}_{\mathrm{EDL}}^{\mathrm{tv}}. \tag{10}$$

*C. Uncertainty-Correctness Alignment Learning*

Based on the Dirichlet distribution in Eq. (3) under the concentration parameters $\boldsymbol{\alpha}_i^{\mathrm{vt}}$, it holds that

$$\mathbb{E}\left(p_j \mid \boldsymbol{\alpha}_i^{\mathrm{vt}}\right) = \hat{p}_{ij}^{\mathrm{vt}} \triangleq \frac{\alpha_{ij}^{\mathrm{vt}}}{S_i^{\mathrm{vt}}}, j = 1,...,K. \tag{11}$$

where $\mathbb{E}(\bullet)$ denotes the expectation under the Dirichlet distribution. From Eq. (11), the predicted correspondence for image $V_i$ is $T_{j^*}$ with $j^* = \arg\max_j \alpha_{ij}^{\mathrm{vt}}$.

However, argmax provides only a point estimate without quantifying retrieval reliability, making it insufficient for selective retrieval. In the vanilla EDL framework [54], the per-query uncertainty $u_i^{vt} = K / S_i^{\mathrm{vt}}$ with $S_i^{\mathrm{vt}} = \sum_j \alpha_{ij}^{\mathrm{vt}}$ reflects only the total amount of evidence, not where it concentrates on the simplex. Two Dirichlet posteriors sharing the same uncertainty can occupy qualitatively different prediction states. As illustrated in Fig. 2 (where the pentagram denotes the correct class), two Dirichlet distributions with the same uncertainty value can exhibit different concentration patterns: the left distribution concentrates near the ground-truth class and supports a reliable prediction, whereas the right distribution is concentrated near the center of the simplex and provides no clear class preference.

To resolve this, we design the UCL loss to learn an uncertainty that is high if the prediction is incorrect, and low if it is correct. In other words, we encourage uncertainty to be aligned with retrieval correctness. This alignment is illustrated in Fig. 3, where lower uncertainty is encouraged for correct decisions (Fig. 3(a)) and higher uncertainty for incorrect decisions (Fig. 3(b)). The UCL loss is expressed as follows:

$$\begin{aligned}\mathcal{L}_{\mathrm{UCL}}^{\mathrm{vt}} = \; & -\mathbb{I}\left(\arg\max_j \hat{p}_{ij}^{\mathrm{vt}} = i\right)\log\left(1-u_i^{\mathrm{vt}}\right) \\ & -\mathbb{I}\left(\arg\max_j \hat{p}_{ij}^{\mathrm{vt}} \neq i\right)\log\left(u_i^{\mathrm{vt}}\right),\end{aligned} \tag{12}$$

where $\mathbb{I}(\cdot)$ is the indicator function. Likewise, we can obtain the UCL loss for text-to-image correspondence, which is given as

$$\begin{aligned}\mathcal{L}_{\mathrm{UCL}}^{\mathrm{tv}} = \; & -\mathbb{I}\left(\arg\max_j \hat{p}_{ij}^{\mathrm{tv}} = i\right)\log\left(1-u_i^{\mathrm{tv}}\right) \\ & -\mathbb{I}\left(\arg\max_j \hat{p}_{ij}^{\mathrm{tv}} \neq i\right)\log\left(u_i^{\mathrm{tv}}\right).\end{aligned} \tag{13}$$

Based on Eqs. (12) and (13), the overall UCL loss can be expressed as

$$\mathcal{L}_{\mathrm{UCL}} = \mathcal{L}_{\mathrm{UCL}}^{\mathrm{vt}} + \mathcal{L}_{\mathrm{UCL}}^{\mathrm{tv}}. \tag{14}$$

By incorporating the UCL loss in Eq. (14) during training, the model is encouraged to assign higher uncertainty to image-to-text correspondences that are difficult to retrieve accurately during testing. This enables the estimated uncertainty to serve as a more reliable indicator for identifying unreliable samples.

*D. Intra-Modal Relationship Learning*

To improve intra-modal representation quality, we introduce intra-modal RL to support EDL. RL leverages pretrained mentor encoders to distill intra-modal similarity structure into the trainable encoders, producing more discriminative image and text features and thereby supporting more informative Dirichlet outputs. This strengthens the basis for aligning uncertainty estimates with retrieval correctness, which is important in CMRSITR because visually similar RS images may share land-cover patterns, object density, or spatial layouts, while semantically related captions may describe the same RS scene at different levels of granularity [65], [66].

We leverage two existing pretrained encoders as mentors, i.e., Unicom [67] for images and GTE (base) [68] for texts, to produce auxiliary intra-modal similarity references [69], [70]. These mentor encoders are used to provide fixed neighborhood structures within each modality, and their parameters are not updated during ELC training. To be more specific, all the image and text samples in the batch are fed into the two existing mentor models, respectively, which yields the corresponding mentor-side image features $\mathbf{v}_i^{\mathrm{m}}$ and text features $\mathbf{t}_i^{\mathrm{m}}$. Next, the self-similarity within each modality is evaluated based on the similarity scores between pairs of image features and pairs of text features obtained from the mentor models, producing $\left\{\langle \mathbf{v}_i^{\mathrm{m}}, \mathbf{v}_j^{\mathrm{m}} \rangle \mid i,j = 1,\ldots,K\right\}$ and $\left\{\langle \mathbf{t}_i^{\mathrm{m}}, \mathbf{t}_j^{\mathrm{m}} \rangle \mid i,j = 1,\ldots,K\right\}$. The self-similarity within each modality from the mentor-side features produces the image-side and text-side intra-modal similarity references. These references are used to guide intra-modal relationship preservation during training and can be precomputed offline.

For the image modality, given any image feature $\mathbf{v}_i^{\mathrm{m}}$ obtained from the mentor model, we define a vector consisting of the similarity scores between $\mathbf{v}_i^{\mathrm{m}}$ and all image features in the batch extracted by the mentor model, which is given as

$$\mathbf{s}_i^{\mathrm{mv}} = \left[ \left\langle \mathbf{v}_i^{\mathrm{m}}, \mathbf{v}_1^{\mathrm{m}} \right\rangle, \left\langle \mathbf{v}_i^{\mathrm{m}}, \mathbf{v}_2^{\mathrm{m}} \right\rangle, \ldots, \left\langle \mathbf{v}_i^{\mathrm{m}}, \mathbf{v}_K^{\mathrm{m}} \right\rangle \right]^{\mathrm{T}}, \forall i. \quad (15)$$

Then, the similarity scores $\mathbf{s}_i^{\mathrm{mv}}$ can be input into the EDL activation function to obtain the mentor-side image evidence vector,

$$\boldsymbol{\alpha}_i^{\mathrm{mv}} = \mathrm{EDL}_{\mathrm{act}}\left(\mathbf{s}_i^{\mathrm{mv}}\right) + \mathbf{1}. \quad (16)$$

Based on Eq. (16), the distribution of the intra-modality relationship among the training samples in the image modality can be represented by $\mathbf{q} = [q_1, \ldots, q_K]^{\mathrm{T}}$ and modeled using a Dirichlet distribution $\mathrm{Dir}\left(\mathbf{q} \mid \boldsymbol{\alpha}_i^{\mathrm{mv}}\right)$. Here, a larger value of $q_j$ indicates that image $V_i$ is more similar to image $V_j$ than other images in the current batch.

Similarly, the student-side intra-modal relationship for image $V_i$ is modeled with a Dirichlet distribution $\mathrm{Dir}\left(\mathbf{q} \mid \boldsymbol{\alpha}_i^{\mathrm{vv}}\right)$. To improve the image encoder of the trained model in terms of feature extraction, knowledge from the mentor model is exploited and transferred to the trained model. This is achieved by encouraging the intra-modal Dirichlet distributions (obtained from both the trained model and the mentor model) to become more similar via the RL loss for RS images,

$$\mathcal{L}_{\mathrm{RL}}^{\mathrm{v}} = \frac{1}{K} \sum_{i=1}^{K} \mathrm{KL}\left[ \mathrm{Dir}\left(\mathbf{q}_i \mid \boldsymbol{\alpha}_i^{\mathrm{vv}}\right) \| \, \mathrm{Dir}\left(\mathbf{q}_i \mid \boldsymbol{\alpha}_i^{\mathrm{mv}}\right) \right]. \quad (17)$$

Minimizing $\mathcal{L}_{\mathrm{RL}}^{\mathrm{v}}$ encourages the trainable image encoder to preserve the mentor-side intra-modal similarity structure, thereby improving image-side representation quality and supporting more informative Dirichlet outputs for EDL. In a similar way, the RL loss $\mathcal{L}_{\mathrm{RL}}^{\mathrm{t}}$ for texts can be calculated. The overall RL loss is given as

$$\mathcal{L}_{\mathrm{RL}} = \mathcal{L}_{\mathrm{RL}}^{\mathrm{v}} + \mathcal{L}_{\mathrm{RL}}^{\mathrm{t}}. \quad (18)$$

Through RL, improved intra-modal features produce more informative Dirichlet evidence at the EDL head.

Finally, the overall training loss can be obtained by combining EDL, UCL, and RL, i.e.,

$$\mathcal{L} = \mathcal{L}_{\mathrm{EDL}} + b_2 \mathcal{L}_{\mathrm{UC}} + b_3 \mathcal{L}_{\mathrm{RL}}, \quad (19)$$

where $b_2$ and $b_3$ are two hyperparameters that can be adjusted to regulate the losses of EDL, UCL, and RL.

### *E. Test Stage*

During the test stage, ELC performs uncertainty-gated retrieval for each incoming query, which can be either an RS image or a text description, as illustrated in Fig. 1. Given a query $Q$, ELC first feeds it into the corresponding encoder to obtain its retrieval embedding. The embedding is then compared with gallery samples from the other modality to compute cross-modal similarity scores. Following the EDL formulation in Section III-B, these scores are passed to the EDL head to obtain the corresponding Dirichlet concentration parameters as in Eq. (3), from which the estimated uncertainty score $u_t$ is computed. The query is compared with an uncertainty threshold $\tau$: if $u_t \le \tau$, the top-$L$ retrieval results are directly returned; otherwise, the query is routed to the RS-TTA stage for refinement. By gating on $u_t$, ELC realizes query-level selective prediction: only queries whose own evidence flags them as unreliable are refined, while confident queries are returned unchanged.

In our experiments, we use a fixed deferral ratio to control the computational budget of RS-TTA. With a fixed deferral ratio $\rho$, the uncertainty threshold $\tau$ is determined as the cutoff that separates the highest-uncertainty fraction of queries from the remaining queries in the current test query set.

For high-uncertainty image queries, two RS-aware operators, *i.e.*, radiometric and geometric operators, are applied. The radiometric operator rescales each color channel as

$$\begin{aligned} I'_c &= g_c I_c + \beta_c, \\ g_c &\sim \mathcal{U}(0.95, 1.05), \\ \beta_c &\sim \mathcal{U}(-0.02, 0.02), \end{aligned} \quad (20)$$

where $c$ indexes the RGB channel, $I_c$ and $I'_c$ are the original and perturbed pixel intensities, $g_c$ is the channel-wise gain, and $\beta_c$ is the channel-wise bias. $\mathcal{U}(\omega_1, \omega_2)$ denotes the uniform distribution on $[\omega_1, \omega_2]$, and $g_c$, $\beta_c$ are sampled independently across channels. This mirrors the multiplicative-gain and additive-bias form used in standard satellite radiometric calibration [39] and reproduces the small inter-acquisition calibration drift observed on operational optical RS sensors; by contrast, generic color jitter introduces color perturbations that are inconsistent with the radiometric physics of RS imagery. The geometric operator rotates the query about its image center by an angle $\theta_r \sim \mathcal{U}(-15^{\circ}, 15^{\circ})$, using bilinear sampling with reflection padding. Small-angle rotation is applied for RS imagery for two complementary reasons. First, satellite acquisitions exhibit substantial in-plane orientation variability due to sensor viewing geometry, flight path, and tile cropping, so the model must be robust to mild orientation shifts of the same scene [71]. Second, RS caption annotations are often defined with respect to the specific orientation of the cropped patch; therefore, large rotations may invalidate spatial-relational descriptions in the ground-truth captions. Restricting rotations to small angles helps preserve

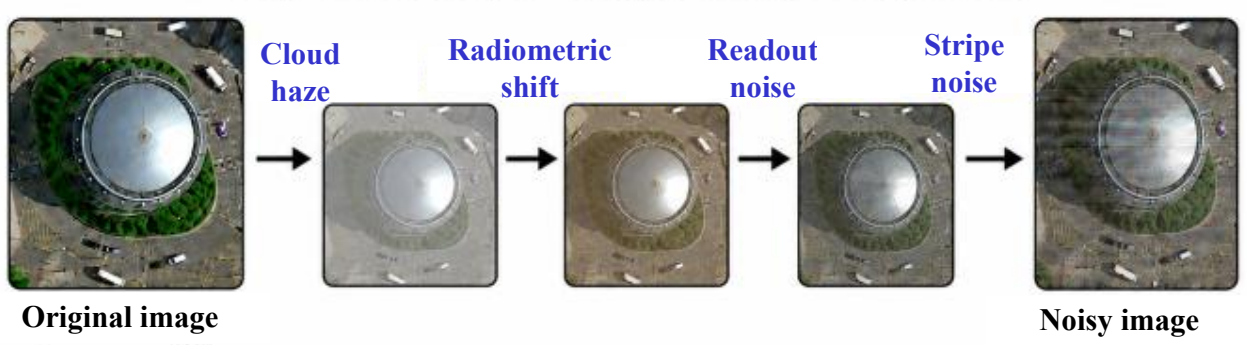

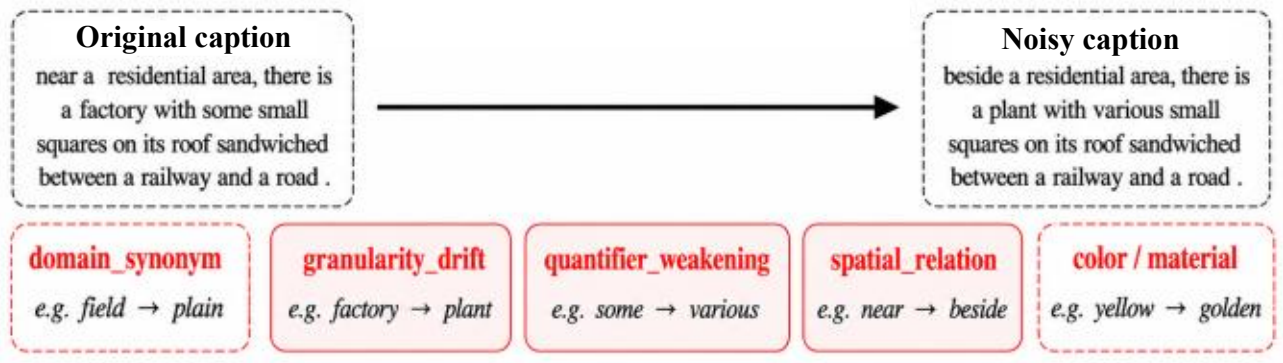

Fig. 4. Illustration of noisy datasets generation process.

caption semantics while still evaluating the orientation invariance expected of the model [72].

For high-uncertainty text queries, RS-vocabulary substitution replaces matched RS-domain head words drawn from a 174-entry curated lexicon, which spans five axes: domain synonym, granularity drift, quantifier weakening, spatial relation, and color/material substitution. We construct a 174-entry curated lexicon from the union of the training-caption vocabularies of the RSICD [73] and the RSITMD [6] datasets, retaining multi-word phrases as atomic entries. At inference, each non-overlapping matched word or phrase in the caption is independently selected for substitution with a per-match probability $p_{\text{sub}}$ , and replaced by a uniformly sampled alternative from the same semantic axis, with at most three substitutions per caption. Unconstrained generic text-augmentation operators like EDA [64] are unsuitable for RS captions because they may map domain-specific terms such as 'harbor' or 'viaduct' to non-RS alternatives and thereby destroy the spatial-semantic content; the empirical gap between the two regimes is quantified in Section IV-G.

For each deferred query, four independently augmented copies are produced. Embeddings of these four augmented queries are averaged together with the original query embedding followed by $l_2$ normalization to form the final retrieval feature, which is then used to recompute the cross-modal ranking. This selective deferral helps stabilize predictions and suppress the impact of noisy inputs in uncertain conditions, thereby improving the overall robustness of CMRSITR. The overall training and test procedure of ELC is summarized in Algorithm 1.

**Algorithm 1** Training and Test Procedure of ELC

***Training stage***

**Input:** Training set $D = \{(V_i, T_i)\}$ of $n$ image-text pairs; trainable image and text encoders; frozen mentor encoders; batch size $K$; number of epochs; loss weights $b_1$, $b_2$, $b_3$.

**Output:** Trained image and text encoders.

1: Precompute the mentor-side intra-modal similarity references (offline).
2: **for** each epoch **do**
3:     **for** each mini-batch of $K$ image–text pairs $(V_i, T_i)$ **do**
4:         Extract image features $\mathbf{v}_i$ and text features $\mathbf{t}_i$ with the trainable encoders.
5:         Compute the inter-modal similarity scores $\mathbf{s}_i^{vt}$ ($\mathbf{s}_i^{tv}$) (Eq. 1).
6:         Obtain the evidence $\mathbf{e}_i^{vt}$ ($\mathbf{e}_i^{tv}$) the Dirichlet parameters $\boldsymbol{\alpha}_i^{vt}$ ($\boldsymbol{\alpha}_i^{tv}$) and the uncertainty $u_i^{vt}$ ($u_i^{tv}$) (Eqs. 2–4).
7:         Compute the EDL loss $L_{\text{EDL}}$ (Eqs. 5–10).
8:         Compute the uncertainty–correctness alignment loss $L_{\text{UCL}}$ (Eqs. 11–14).
9:         Compute the intra-modal relationship learning loss $L_{\text{RL}}$ (Eqs. 15–18).
10:        Update the trainable encoders with the total loss $L = L_{\text{EDL}} + b_2 L_{\text{UCL}} + b_3 L_{\text{RL}}$ (Eq. 19).
11:     **end for**
12: **end for**

***Test stage (uncertainty-gated selective retrieval)***

**Input:** Query $Q$ (RS image or text); gallery of the other modality; trained encoders; deferral ratio $\rho$; return size $L$.

**Output:** Top-$L$ retrieval list for $Q$.

13: Encode $Q$ and compute its cross-modal similarity scores to the gallery.
14: Obtain the Dirichlet parameters (Eq. 3) and the uncertainty $u_t$ of $Q$.
15: Set the threshold $\tau$ to defer the most-uncertain fraction $\rho$ of the queries.
16: **if** $u_t \le \tau$ **then**
17:     Return the top-$L$ retrieval results directly.
18: **else**
19:     Refine $Q$ with RS-TTA, then return the top-$L$ retrieval results.
20: **end if**

## IV. Experimental Results

### *A. Dataset and Evaluation Metrics*

To validate the effectiveness of the proposed ELC method, we conduct experiments on two widely used datasets, i.e., RSICD [73] and RSITMD [6]. RSICD [73] contains 10,921 RS images collected from Google Earth, each paired with five manually annotated textual descriptions. RSITMD [6] consists of 4,743 RS images, where each image is associated with five fine-grained textual descriptions. Following the previous work [9], all the datasets are divided into 80% for training, 10% for testing, and 10% for validation.

To better simulate real-world conditions, we further construct noisy test datasets based on the original RSICD and RSITMD datasets, referred to as Noisy RSICD-Test and Noisy RSITMD-Test, respectively, where sensor- and atmosphere-physics (cloud/haze, radiometric drift, readout, push-broom striping) and caption annotation heterogeneity are considered, as shown in Fig. 4.

For the RS image modality, we apply four perturbations following the optical-RS image formation and sensing chain. Spatially varying cloud and haze attenuation follows the Koschmieder model [37]. Let $\mathbf{J}(x) \in [0,1]^3$ denote the original RGB value at pixel location $x$, and let $\tilde{\mathbf{I}}_{ch}$ denote the perturbed value. Cloud/haze attenuation is generated using the atmospheric scattering form

$$\tilde{\mathbf{I}}_{ch}(x) = t(x)\mathbf{J}(x) + (1 - t(x))A, \tag{21}$$

where $t(x) \in [0,1]$ is the pixel-wise transmission map and $A$ is the airlight intensity. $t(x)$ is sampled from a random field smoothed by a Gaussian kernel [38]. Next, based on standard radiometric calibration and rescaling formulations for optical satellite data [39], we simulate radiometric variation through the per-channel affine transform:

$$\tilde{\mathbf{I}}_r(x) = g_c \tilde{\mathbf{I}}_{ch}(x) + \beta_c, \tag{22}$$

with notation consistent with Eq. (20). We then add zero-mean readout noise [40] with standard deviation $\sigma_{\text{readout}}$,

$$\tilde{\mathbf{I}}_{rd}(x) = \tilde{\mathbf{I}}_r(x) + \epsilon_{rd}(x), \quad \epsilon_{rd}(x) \sim \mathcal{N}(0, \sigma^2_{\text{readout}}), \tag{23}$$

where $\epsilon_c(x)$ is independently sampled for each pixel and channel, and $\sigma_{\text{readout}}$ controls its standard deviation. Finally, row-wise stripe noise with standard deviation $\sigma_{\text{stripe}}$ is added, where the stripe offset is shared along each image row to mimic push-broom striping effects.

$$\tilde{\mathbf{I}}_s(x) = \tilde{\mathbf{I}}_{rd}(x) + \epsilon_{col}, \quad \epsilon_{col} \sim \mathcal{N}(0, \sigma^2_{\text{stripe}}), \tag{24}$$

where $\epsilon_{col}$ is a row-dependent offset shared along all columns of row $col$, following the structured form of push-broom striping artifacts [41]. The final perturbed image is obtained by clipping $\tilde{\mathbf{I}}_s$ to $[0,1]$. The perturbation parameters are set as $A = 0.85$, $t(x) \in [0.55, 0.80]$, $g_c \in [0.90, 1.10]$, $b_c \in [-0.03, 0.03]$, $\sigma_{\text{readout}} = 0.02$, and $\sigma_{\text{stripe}} = 0.025$. These parameters produce visible degradation while preserving the scene semantics.

For the text modality, we replace RS-domain head words using a curated 174-entry dictionary spanning five lexical axes: domain synonym, granularity drift, quantifier weakening, spatial relation, and color/material substitution. For each caption, matches are resolved longest-first, and each non-overlapping match is independently replaced with probability $p_{\text{sub}} = 0.30$, where $p_{\text{sub}}$ denotes the per-match substitution probability. To prevent over-corruption of short captions, the number of substitutions per caption is capped at 3. Unlike generic synonym replacement, this substitution is operationally meaningful for RS, e.g., 'harbor' may map to 'dock' or 'marina', but never to 'desk'.

For each dataset, we form a 50/50 mixed test split by randomly assigning half of the test images and their associated captions to the noisy stream, while keeping the remaining half clean. Within the noisy stream, every image is perturbed by the image-side protocol, while only captions containing matched RS-domain head words are modified. The 50/50 split simulates deployment where degraded and clean queries co-exist. For both retrieval directions, the resulting 50/50 split is used as both the query set and the gallery.

To assess the retrieval performance, we adopt the widely used Recall at $L$ (R@$L$) metric with $L \in \{1,5,10\}$. R@$L$ measures the percentage of queries whose corresponding ground-truth matches appear within the top-$L$ retrieved results. Formally, it is defined as

$$R@L = \frac{1}{N}\sum_{i=1}^{N} \mathbb{I}(\text{rank}(i) \le L), \tag{25}$$

where $N$ denotes the total number of queries, rank($i$) represents the ranking position of the correct match for the $i$-th query. A higher R@$L$ value indicates higher retrieval accuracy. In addition, we also report the Recall Sum (RSUM), which provides an overall evaluation of retrieval performance across multiple thresholds and both retrieval directions (image-to-text and text-to-image). It is defined as

$$RSUM = \sum_{L \in \{1,5,10\}} \left( R^{i2t}@L + R^{t2i}@L \right), \tag{26}$$

where $R^{i2t}@L$ and $R^{t2i}@L$ denote the recall scores for image-to-text and text-to-image retrieval, respectively.

### *B. Implementation Details*

We implement the experiments using the PyTorch framework on one NVIDIA A100 with 80 GB of memory. The backbone of our model follows the CLIP ViT-B/32 architecture [74], where the image encoder and text encoder are initialized from the pretrained CLIP model and subsequently finetuned on the target datasets to adapt to the RS domain. Following Section III-D, Unicom [67] and GTE (base) [68] are adopted as frozen image and text mentor encoders to generate intra-modal similarity references for RL. These mentors are used only as fixed intra-modal similarity references. The length of the retrieval feature is set to 512. All the methods are trained by the AdamW [75] optimizer with a learning rate of $10^{-6}$, and a batch size of 128 for 40 epochs. The hyperparameters $b_1$, $b_2$, and $b_3$ are set to 40, 1, and 1. The weight decay is set to 0.1, and a cosine annealing learning rate schedule is applied to ensure smooth convergence.

The base version of our method without RS-TTA is denoted as ELC0, while the full model ELC applies our RS-TTA at inference. We fix the deferral ratio $\rho = 0.1$. For each retrieval direction, the test queries are ranked by estimated uncertainty and the most-uncertain 10% are deferred to RS-TTA, while the remaining queries are returned directly. For each deferred query, four augmented versions are generated using three RS-aware operators: (i) radiometric perturbation; (ii) random small-angle rotation within ±15°; and (iii) RS-vocabulary substitution drawn from the 174-entry lexicon defined in Section IV-A. The embeddings of the four augmented versions are averaged with the original embedding to form the final retrieval feature. We discuss the role of the 10% deferral ratio in Section IV-F, and contrast RS-TTA with generic TTA baseline in Section IV-G.

### *C. Comparisons with State-of-the-Art Methods*

In this subsection, we present a comprehensive comparison between our proposed method and the state-of-the-art approaches. We compare our proposed method with 17 state-of-the-art methods, including 4 methods using traditional convolutional neural networks and recurrent neural networks as backbones: AMFMN [6], GaLR [9], SMLGN [34] and KAMCL [22], and 13 methods with CLIP backbones (ViT-B/32), including CLIP [74], SIRS [35], CUP [27], VSE++ [76], Poly [77], VSE∞ [78], InfoNCE [79], PAU [80], AIR [81], PE-RSITR [32], GLGSA [16], FGVLA [21] and RemoteCLIP [25], on both the RSICD and RSITMD datasets. For CLIP-based methods, both the image and text encoders are initialized from the pretrained CLIP model and subsequently finetuned on the target RSICD or RSITMD dataset, except for RemoteCLIP [25], which performs continual pretraining on a large-scale RS dataset containing 165,745 images. All the baseline implementations follow their original settings, using either publicly released source code or pretrained model weights provided by the authors to ensure experimental consistency and fairness.

TABLE I
QUANTITATIVE EVALUATION RESULTS OF IMAGE-TO-TEXT AND TEXT-TO-IMAGE RETRIEVAL ON THE ORIGINAL RSICD AND RSITMD TEST DATASETS. THE BEST RESULTS ARE HIGHLIGHTED IN BOLD AND UNDERLINED. THE SECOND-BEST RESULTS ARE UNDERLINED.

| Method | RSICD | | | | | | | RSITMD | | | | | | |
|---|---|---|---|---|---|---|---|---|---|---|---|---|---|---|
| | Image to Text | | | Text to Image | | | RSUM | Image to Text | | | Text to Image | | | RSUM |
| | R@1 | R@5 | R@10 | R@1 | R@5 | R@10 | | R@1 | R@5 | R@10 | R@1 | R@5 | R@10 | |
| AMFMN [6] | 5.39 | 15.08 | 23.40 | 4.90 | 18.28 | 31.44 | 98.49 | 10.63 | 24.78 | 41.81 | 11.51 | 34.69 | 54.87 | 178.29 |
| GaLR [9] | 6.59 | 19.85 | 31.04 | 4.69 | 19.48 | 32.13 | 113.78 | 14.82 | 31.64 | 42.48 | 11.15 | 36.68 | 51.68 | 188.45 |
| SMLGN [34] | 8.87 | 25.53 | 37.24 | 7.85 | 27.14 | 42.58 | 149.21 | 17.26 | 39.38 | 51.55 | 13.19 | 43.94 | 60.40 | 225.72 |
| KAMCL [22] | 11.99 | 27.17 | 38.33 | 7.78 | 25.23 | 40.02 | 150.52 | 16.81 | 34.96 | 47.12 | 14.60 | 41.86 | 59.86 | 215.21 |
| CLIP [74] | 5.03 | 14.64 | 23.51 | 5.47 | 17.53 | 27.78 | 93.96 | 9.51 | 23.01 | 32.96 | 8.81 | 28.01 | 43.10 | 145.40 |
| SIRS [35] | 10.34 | 27.69 | 39.66 | 8.74 | 30.38 | 49.50 | 166.31 | 18.45 | 44.06 | 57.73 | 19.44 | 43.18 | 58.40 | 241.26 |
| CUP [27] | 12.05 | 32.66 | 48.70 | 9.86 | 31.60 | 47.15 | 182.02 | 21.57 | 43.70 | 56.75 | 17.70 | 48.40 | 64.82 | 252.94 |
| VSE++ [76] | 16.02 | 35.04 | 47.58 | 10.74 | 30.41 | 45.56 | 185.35 | 19.69 | 38.50 | 52.21 | 14.47 | 41.86 | 56.81 | 223.54 |
| Poly [77] | 15.55 | 32.02 | 45.20 | 10.48 | 31.75 | 46.18 | 181.18 | 21.02 | 35.84 | 47.79 | 13.98 | 41.90 | 60.09 | 220.62 |
| VSE∞ [78] | 16.10 | 36.51 | 48.76 | 11.18 | 32.83 | 47.70 | 193.08 | 21.02 | 46.02 | 57.74 | 16.73 | 47.21 | 61.37 | 250.09 |
| InfoNCE [79] | 15.28 | 33.39 | 47.48 | 11.89 | 35.31 | 53.14 | 196.49 | 19.47 | 39.60 | 53.32 | 17.26 | 47.88 | 67.61 | 245.14 |
| PAU [80] | 15.10 | 34.31 | 48.12 | 12.39 | 35.00 | 53.47 | 198.39 | 22.12 | 41.37 | 54.87 | 16.37 | 49.20 | 66.46 | 250.39 |
| AIR [81] | 16.29 | 33.58 | 46.39 | 12.19 | 37.07 | 53.94 | 199.46 | 24.12 | 42.04 | 54.20 | 19.20 | 50.22 | 68.89 | 258.67 |
| PE-RSITR [32] | 14.13 | 31.51 | 44.78 | 11.63 | 33.92 | 50.73 | 186.70 | 23.67 | 44.07 | 60.36 | 20.10 | 50.63 | 67.97 | 266.80 |
| GLGSA [16] | 14.38 | 36.73 | 45.35 | 11.75 | 34.07 | 50.12 | 192.40 | 26.77 | 48.23 | 60.39 | 22.52 | 52.39 | 69.46 | 279.76 |
| FGVLA [21] | 16.93 | 36.23 | 48.49 | 12.72 | 36.61 | 53.76 | 204.74 | 26.33 | **51.99** | 64.38 | 21.77 | 54.12 | 72.92 | 291.51 |
| RemoteCLIP [25] | 17.02 | 37.97 | 51.51 | 13.71 | 37.11 | 54.25 | 211.57 | **27.88** | 50.66 | **65.71** | 22.17 | **56.46** | **73.41** | **296.29** |
| **ELC0** | 17.93 | 37.88 | **51.88** | **14.03** | 37.37 | 54.38 | 213.47 | 27.21 | 50.22 | 63.27 | 22.65 | 55.97 | 72.52 | 291.84 |
| **ELC** | **18.12** | **38.55** | 51.78 | **14.03** | **38.01** | **54.71** | **215.20** | 27.43 | 51.11 | 63.72 | **23.32** | 56.06 | 72.43 | 294.07 |

| Query | | Top 5 Retrieval Results (RSICD) | Query | | Top 5 Retrieval Results (RSITMD) |
|---|---|---|---|---|---|
| the round white building and a square building are enclosed by a circular parking lot and a circular road . | Remote CLIP | | On the apron surrounded by the airport runway, there is a Y-shaped terminal, and some planes are parked orderly. | Remote CLIP | |
| | ELC | | | ELC | |
| | Remote CLIP | 1. there is a zigzag building with a swimming pool in the resort next to which is a parking lot and two tennis fields.<br>2. we can see that a unique zigzag building a swimming pool a parking lot and two tennis fields are in the resort.<br>3. we can see that a unique zigzag building a swimming pool a parking lot and two tennis fields are in the resort.<br>4. two yellow tennis courts sits next to this smart white building in this resort surrounded by sand .<br>5. several tennis courts and a swimming pool can be fond in this resort. | | Remote CLIP | 1. Green grass revolves around a large gray stadium.<br>2. Green grass revolves around a large gray stadium.<br>3. Green grass surrounds a large gray stadium.<br>4. Green grass surrounds a big gray stadium.<br>5. three sides of blearchers belonging to the football fields have ceilings and the other side has three layers of blearchers without ceiling. |
| | ELC | 1.we can see that a unique zigzag building a swimming pool a parking lot and two tennis fields are in the resort.<br>2.we can see that a unique zigzag building a swimming pool a parking lot and two tennis fields are in the resort.<br>3.there is a zigzag building with a swimming pool in the resort next to which is a parking lot and two tennis fields.<br>4.a large building with cars and two tennis courts are in a resort with some tropical green trees.<br>5.two yellow tennis courts sits next to this smart white building in this resort surrounded by sand. | | ELC | 1. Green grass revolves around a large gray stadium.<br>2. Green grass revolves around a large gray stadium.<br>3. Green grass surrounds a large gray stadium.<br>4. a lawn and a runway are in the stadium.<br>5. Green grass surrounds a big gray stadium. |

Fig. 5. Query examples with the corresponding top-5 retrieved results on the RSICD and RSITMD datasets. For image retrieval, samples enclosed in green boxes denote correctly retrieved images, while those in red boxes represent false positives. For text retrieval, correctly retrieved sentences are shown in green text, whereas incorrectly retrieved ones are highlighted in red text.

TABLE II
QUANTITATIVE EVALUATION RESULTS OF IMAGE-TO-TEXT AND TEXT-TO-IMAGE RETRIEVAL ON THE NOISY RSICD-TEST AND NOISY RSITMD-TEST DATASETS. THE BEST RESULTS ARE HIGHLIGHTED IN BOLD AND UNDERLINED. THE SECOND-BEST RESULTS ARE UNDERLINED.

| Method | Noisy RSICD-Test | | | | | | | Noisy RSITMD-Test | | | | | | |
|---|---|---|---|---|---|---|---|---|---|---|---|---|---|---|
| | Image to Text | | | Text to Image | | | RSUM | Image to Text | | | Text to Image | | | RSUM |
| | R@1 | R@5 | R@10 | R@1 | R@5 | R@10 | | R@1 | R@5 | R@10 | R@1 | R@5 | R@10 | |
| CLIP [74] | 4.21 | 12.81 | 20.13 | 3.93 | 14.68 | 23.64 | 79.40 | 7.08 | 16.81 | 25.66 | 6.90 | 22.83 | 36.90 | 116.19 |
| VSE++ [76] | 13.27 | 29.19 | 40.81 | 9.30 | 26.13 | 39.01 | 157.71 | 14.60 | 32.74 | 45.35 | 12.35 | 34.51 | 49.73 | 189.29 |
| Poly [77] | 12.17 | 28.18 | 39.25 | 8.34 | 25.34 | 38.72 | 152.00 | 15.27 | 31.19 | 41.81 | 11.28 | 35.84 | 52.61 | 188.01 |
| VSE∞ [78] | 12.26 | **31.38** | 43.00 | 9.46 | 27.98 | 41.12 | 165.20 | 16.81 | 36.50 | 48.67 | 13.54 | 41.50 | 58.23 | 215.27 |
| PAU [80] | 11.62 | 26.26 | 38.43 | 9.17 | 28.91 | 43.59 | 157.97 | 16.37 | 34.51 | 46.02 | 15.04 | 42.70 | 59.60 | 214.25 |
| InfoNCE [79] | 12.17 | 28.09 | 41.99 | 9.57 | 27.98 | 42.20 | 161.99 | 16.59 | 35.62 | 48.01 | 14.38 | 41.95 | 60.18 | 216.73 |
| AIR [81] | 10.61 | 28.82 | 41.45 | 9.55 | 26.57 | 39.40 | 156.40 | 21.46 | 36.95 | 48.01 | 15.88 | 43.72 | 62.17 | 228.19 |
| PE-RSITR [32] | 14.00 | 29.46 | 39.34 | 7.83 | 23.15 | 33.52 | 147.30 | 17.70 | 33.63 | 45.58 | 12.26 | 36.24 | 50.62 | 196.02 |
| FGVLA [21] | 12.17 | 29.09 | 40.81 | **11.00** | 30.54 | 46.02 | 169.62 | 16.59 | 42.26 | 56.42 | 18.10 | 44.87 | 60.27 | 238.50 |
| RemoteCLIP [25] | 11.99 | 28.82 | 41.17 | 9.94 | 30.72 | **47.08** | 169.72 | 21.24 | 42.26 | 53.98 | 17.92 | 44.42 | 60.31 | 240.13 |
| **ELC0** | 14.82 | 31.02 | 43.66 | 10.44 | 31.16 | 44.36 | 175.46 | 21.86 | 42.49 | 56.42 | 18.06 | **48.67** | 63.90 | 251.40 |
| **ELC** | **15.28** | 30.56 | **45.38** | 10.83 | **31.24** | 45.64 | **178.93** | **23.01** | **44.03** | **57.30** | **19.03** | 47.60 | **65.67** | **256.64** |

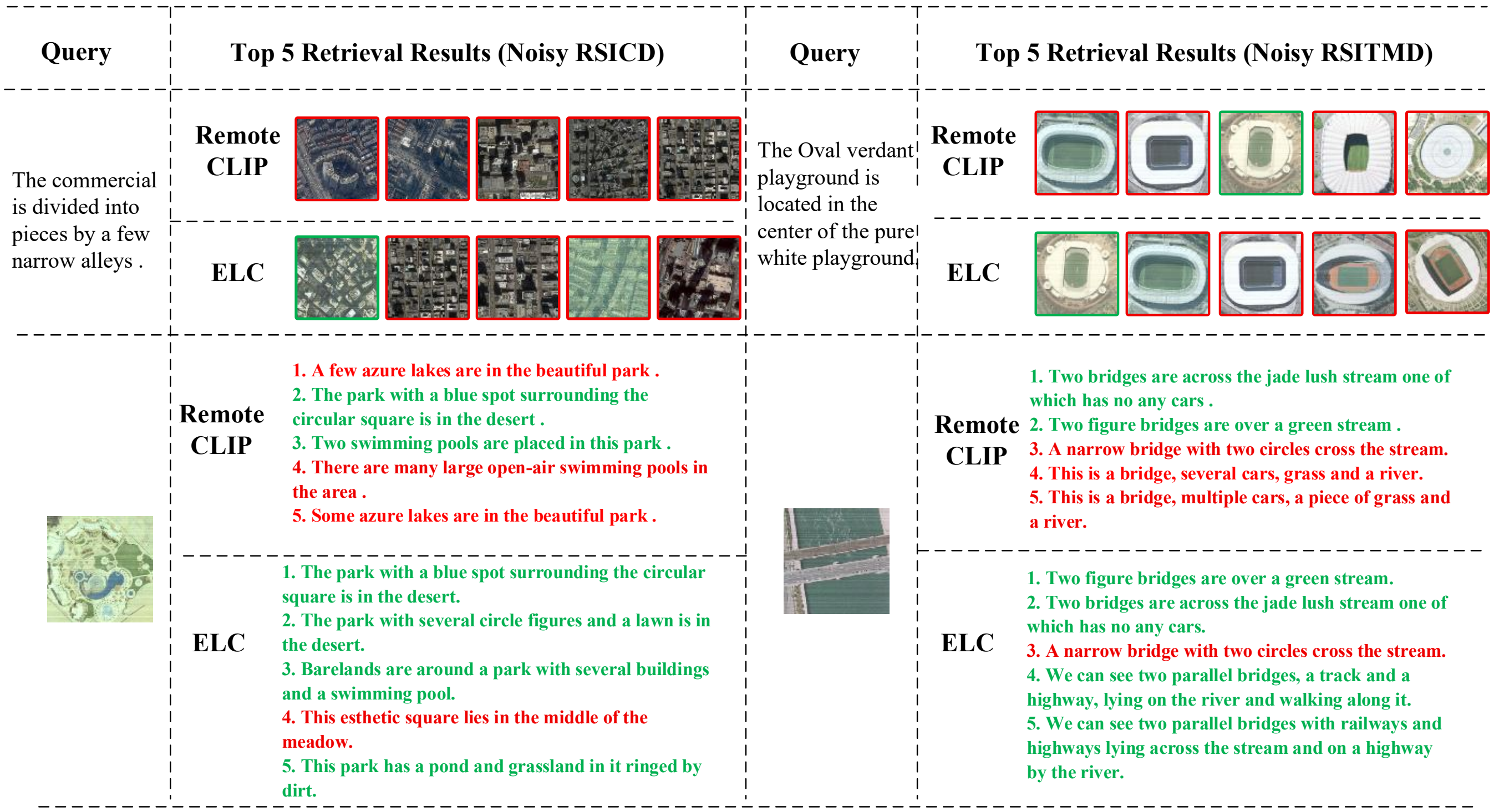

Fig. 6. Query examples with the corresponding top-5 retrieved results on the Noisy RSICD-Test and Noisy RSITMD-Test datasets. For image retrieval, samples enclosed in green boxes denote correctly retrieved images, while those in red boxes represent false positives. For text retrieval, correctly retrieved sentences are shown in green text, whereas incorrectly retrieved ones are highlighted in red text.

**Results on RSICD**. As shown in Table I, the proposed ELC0 consistently achieves the highest overall retrieval accuracy among all the existing state-of-the-art methods (except for the image-to-text retrieval metric R@5), even surpassing RemoteCLIP [25], which is trained on a dataset several times larger than RSICD. In terms of overall performance, the RSUM metric improves from 211.57 to 213.47, confirming the effectiveness of the proposed method in enhancing retrieval accuracy. Furthermore, when RS-TTA is introduced, ELC achieves an additional improvement, reaching an RSUM of 215.20. This demonstrates that uncertainty-guided test-time refinement effectively improves retrieval accuracy by mitigating the influence of high-uncertainty predictions. In Fig. 5, we present a visual comparison of the top-5 retrieval results obtained by the proposed ELC and the second-best method, RemoteCLIP. For the text-to-image retrieval task, RemoteCLIP fails to return the correct image within the top five results, whereas the proposed method successfully retrieves it at the third position. For the image-to-text retrieval task, the samples retrieved by ELC exhibit strong semantic consistency with the query, with all top-5 retrieved sentences being correct. In contrast, the competing method [25] also retrieves mostly relevant samples but includes one incorrect sentence among the top five results.

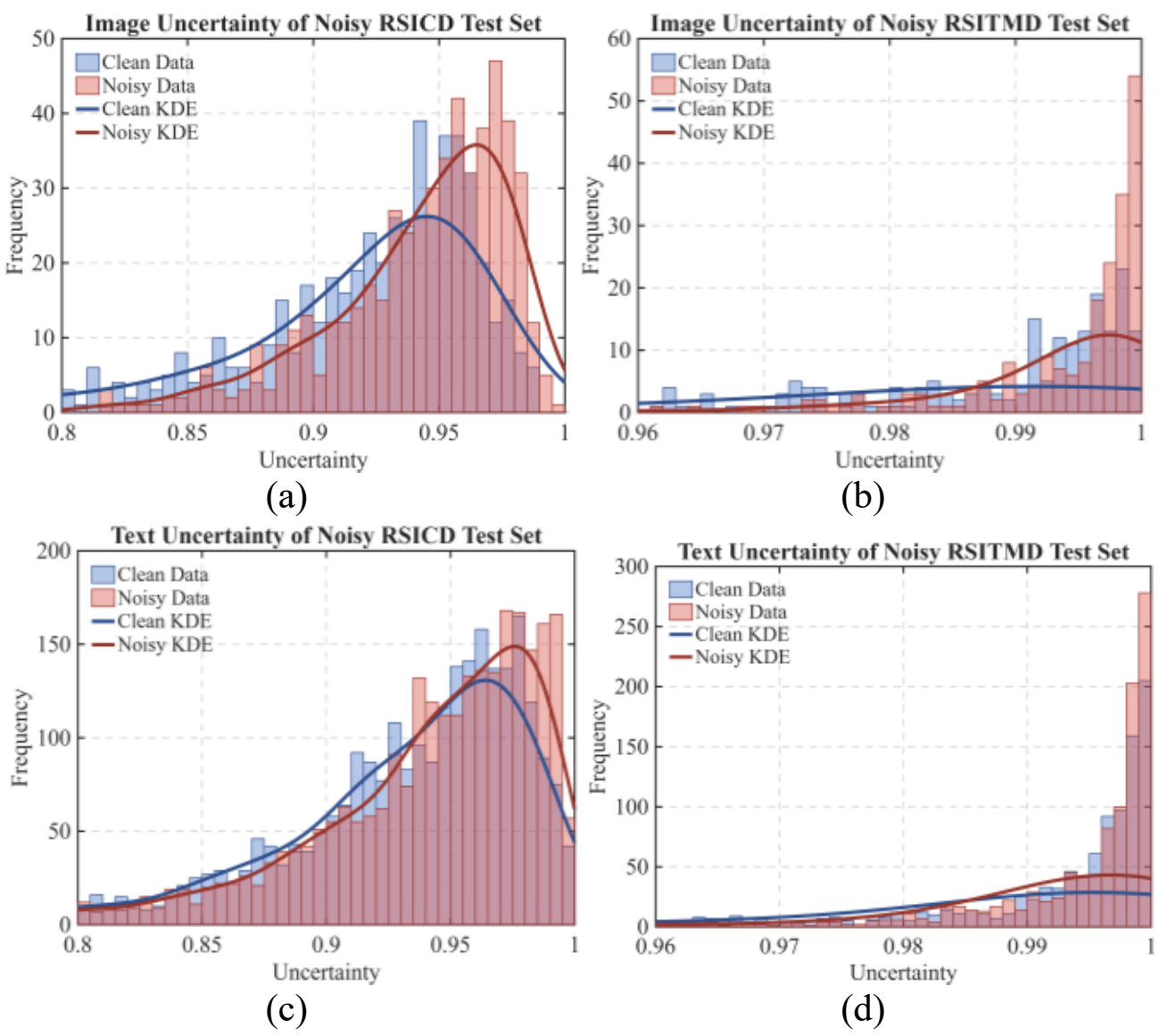

Fig. 7. Analysis of uncertainty distributions on noisy datasets. (a) Image uncertainty on Noisy RSICD-Test dataset. (b) Image uncertainty on Noisy RSITMD-Test dataset. (c) Text uncertainty on Noisy RSICD-Test dataset. (d) Text uncertainty on Noisy RSITMD-Test dataset.

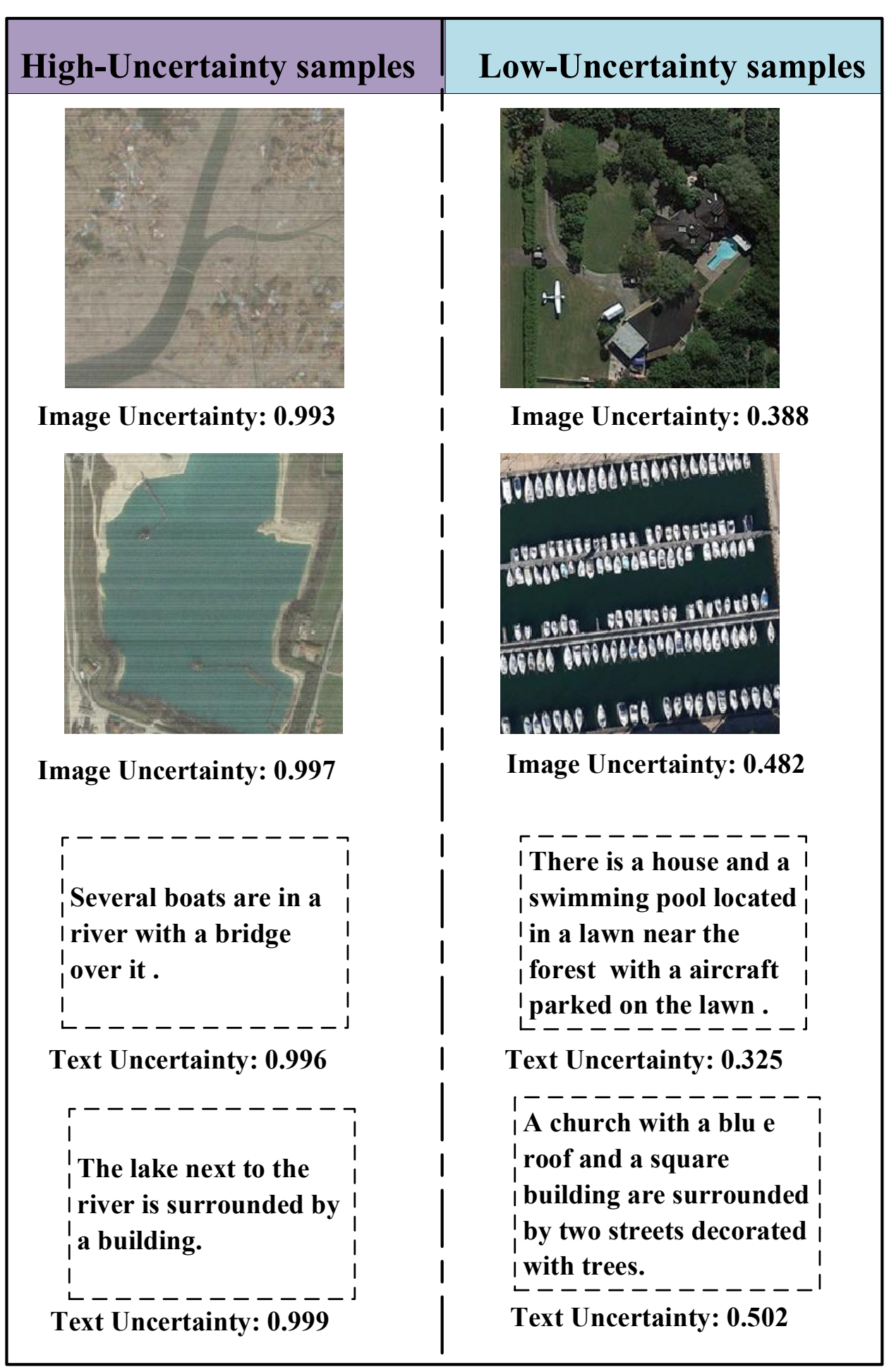

Fig. 8. Representative low- and high-uncertainty image and text samples from the Noisy RSICD-Test and RSITMD-Test datasets.

**Results on RSITMD**. The right half of Table I presents the retrieval performance on the original RSITMD dataset, which features more fine-grained and semantically detailed textual annotations. The proposed ELC0 achieves the third-best overall performance in terms of RSUM, ranking closely behind RemoteCLIP [25]. The refined version, ELC, further enhances retrieval accuracy through uncertainty-guided test-time refinement, achieving the second-highest overall retrieval performance after RemoteCLIP. Although RemoteCLIP [25] benefits from continual pretraining on a substantially larger RS dataset, our method attains comparable results despite being trained solely on the much smaller RSITMD dataset, demonstrating its strong generalization under limited data conditions. As shown in Fig. 5, on the RSITMD dataset, the proposed ELC achieves retrieval results comparable to RemoteCLIP. For the text-to-image task, ELC retrieves the correct image at the second position, whereas RemoteCLIP ranks it first. For the image-to-text task, our proposed method successfully retrieves three correct sentences within the top-5 results, compared to four for RemoteCLIP.

**Results on Noisy RSICD-Test dataset**: The left half of Table II presents the performance on the Noisy RSICD-Test dataset. Our base model, ELC0, surpasses all competing methods, including RemoteCLIP [25], achieving an RSUM score of 175.46 compared with 169.72 for RemoteCLIP. Furthermore, when RS-TTA is applied, the performance of ELC0 is further improved. Compared with the second-best method (RemoteCLIP [25]), ELC achieves a 9.21-point improvement, confirming the strong capability of our method to handle noisy inputs. Compared with the results on the original RSICD test dataset, the advantage of the proposed ELC becomes more pronounced with data degradation, while the inclusion of RS-TTA further enhances its robustness. As illustrated by the query examples in Fig. 6, the proposed ELC demonstrates more stable retrieval behavior under noisy conditions. For the text-to-image retrieval task, ELC successfully retrieves the correct image at the first position, whereas all five RemoteCLIP [25] results are mismatched. For the image-to-text retrieval task, ELC retrieves four semantically consistent sentences among the top-5 results, while RemoteCLIP [25] retrieves two correct sentences.

**Results on Noisy RSITMD-Test dataset**: As shown in Table II, the proposed ELC0 obtains an RSUM score of 251.40 on the Noisy RSITMD-Test dataset, outperforming all competing approaches, including the best-performing method on the original RSITMD test dataset, RemoteCLIP [25]. After integrating RS-TTA, ELC achieves the higher RSUM score of 256.64. The proposed ELC achieves superior retrieval performance under noisy conditions, with its advantage becoming increasingly evident compared with the results obtained on the original RSITMD test dataset. As shown in the query examples of Fig. 6, the proposed ELC achieves more accurate and stable retrieval on the Noisy RSITMD-Test dataset, ranking target images higher and retrieving more semantically correct sentences than RemoteCLIP [25] under noisy conditions.

TABLE III
QUANTITATIVE EVALUATION RESULTS OF ABLATION STUDY ON THE RSICD DATASET.

| $\mathcal{L}_{EDL}$ | UW | $\mathcal{L}_{RL}$ | $\mathcal{L}_{UCL}$ | RSICD | | | | | | |
|---|---|---|---|---|---|---|---|---|---|---|
| | | | | Image to Text | | | Text to Image | | | RSUM |
| | | | | R@1 | R@5 | R@10 | R@1 | R@5 | R@10 | |
| √ | | | | 15.83 | 35.68 | 49.59 | 11.89 | 36.72 | 52.94 | 202.65 |
| √ | √ | | | 17.38 | 36.51 | 49.31 | 12.77 | 36.85 | 52.10 | 204.92 |
| √ | √ | √ | | 17.75 | **38.06** | 51.14 | 13.10 | 37.73 | 52.75 | 210.54 |
| √ | √ | | √ | 17.38 | 37.05 | 50.05 | 12.85 | **38.19** | 54.33 | 209.84 |
| √ | √ | √ | √ | **17.93** | 37.88 | **51.88** | **14.03** | 37.37 | **54.38** | **213.47** |

TABLE IV
QUANTITATIVE EVALUATION RESULTS OF ABLATION STUDY ON THE RSITMD DATASET.

| $\mathcal{L}_{EDL}$ | UW | $\mathcal{L}_{RL}$ | $\mathcal{L}_{UCL}$ | RSITMD | | | | | | |
|---|---|---|---|---|---|---|---|---|---|---|
| | | | | Image to Text | | | Text to Image | | | RSUM |
| | | | | R@1 | R@5 | R@10 | R@1 | R@5 | R@10 | |
| √ | | | | 23.89 | 46.02 | 60.84 | 21.64 | 52.88 | 70.44 | 275.71 |
| √ | √ | | | 23.67 | 49.12 | 61.28 | 21.86 | 53.81 | 70.53 | 280.27 |
| √ | √ | √ | | 25.88 | 48.67 | **63.50** | **22.74** | 54.87 | 72.08 | 287.74 |
| √ | √ | | √ | 24.78 | 49.56 | 62.17 | 22.48 | 55.18 | 71.33 | 285.49 |
| √ | √ | √ | √ | **27.21** | **50.22** | 63.27 | 22.65 | **55.97** | **72.52** | **291.84** |

### *D. Effectiveness of Uncertainty Estimation*

To verify the reliability of the uncertainty estimated by our proposed method, we conduct a detailed distribution analysis.

Figure 7 reports per-sample uncertainty histograms together with their kernel density estimates (KDE) for the Noisy RSICD-Test and Noisy RSITMD-Test datasets. As observed, the noisy samples (in red) exhibit higher uncertainty values than the clean samples (in blue) across both datasets, for both image and text. This distributional shift demonstrates that the uncertainty learned by the proposed ELC0 correlates with data quality, enabling the model to identify unreliable samples within the dataset. This capability provides a basis for subsequent uncertainty-guided refinement through RS-TTA, which can further improve the robustness and reliability of CMRSITR systems.

To further illustrate the interpretability of the estimated uncertainty, Fig. 8 presents representative low- and high-uncertainty image and text samples from the Noisy RSICD-Test and RSITMD-Test datasets. For images, high uncertainty is mainly associated with weak semantic distinctiveness along with noise perturbation. The high-uncertainty samples tend to contain repetitive textures, limited structural diversity, or weakly salient scene elements, which reduce the discriminative evidence available for cross-modal matching. By contrast, the low-uncertainty samples contain clearer semantic landmarks and more recognizable spatial organization, such as aircraft and densely arranged boats. For texts, the high-uncertainty samples are generally short and weakly informative, providing only coarse scene-level descriptions with limited discriminative cues. In contrast, the low-uncertainty captions include more specific object categories, attributes, and spatial relations. These examples provide intuitive evidence that the uncertainty estimated by the proposed method is closely related to the visual clarity of image samples and semantic interpretability of text samples.

### *E. Ablation Study*

To validate that each of the three training losses contributes to our proposed ELC, we conduct ablation studies on RSICD and RSITMD, progressively enabling the EDL loss with UW, the UCL loss, and the RL loss. Quantitative results are reported in Tables III and IV.

As shown in Tables III and IV, similar trends can be observed on both datasets. Starting from the baseline model using only $\mathcal{L}_{EDL}$, introducing UW consistently improves the RSUM, showing that UW is beneficial in the training phase by reducing the adverse impact of ambiguous correspondences. Further incorporating $\mathcal{L}_{UCL}$ leads to a clear performance gain on both datasets, showing that UCL contributes effectively to retrieval performance by uncertainty–correctness alignment. Adding $\mathcal{L}_{RL}$ also brings consistent improvement over the baseline, confirming that RL effectively enhances cross-modal discrimination by preserving intra-modal structural relationships.

When all the components are jointly adopted, the full model achieves the best results on both datasets, reaching RSUM values of 213.47 on RSICD and 291.84 on RSITMD. Compared with the baseline, the full model improves the RSUM by 10.82 and 16.13 on RSICD and RSITMD, respectively. These results demonstrate that UW, $\mathcal{L}_{UCL}$ and $\mathcal{L}_{RL}$ are all effective.

### *F. Hyperparameter Sensitivity Analysis*

We analyze the sensitivity of ELC to the loss-balancing hyperparameters $\{b_1, b_2, b_3\}$ by varying one coefficient at a time while keeping the others fixed at the setting used in Section IV-B. As shown in Fig. 9 and Fig. 10, the retrieval performance changes smoothly around the selected configuration and remains stable under moderate variations of each coefficient. This observation indicates that ELC does not rely on a narrowly tuned hyperparameter setting, and that the adopted configuration provides a stable balance among the EDL, UCL, and RL objectives.

We further analyze the effect of the deferral ratio $\rho$ on retrieval performance over the Noisy RSICD-Test and Noisy

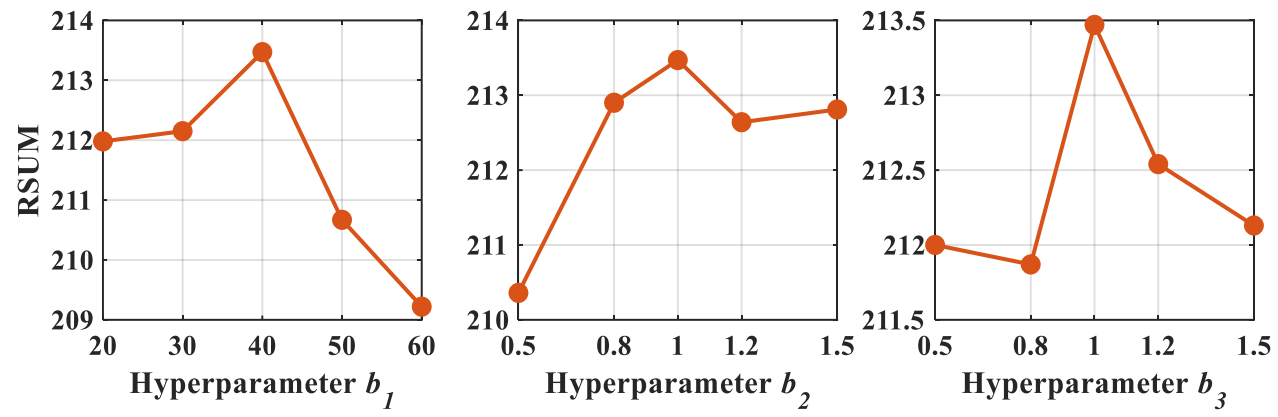

Fig. 9. Sensitivity analysis of hyperparameters on the RSICD dataset.

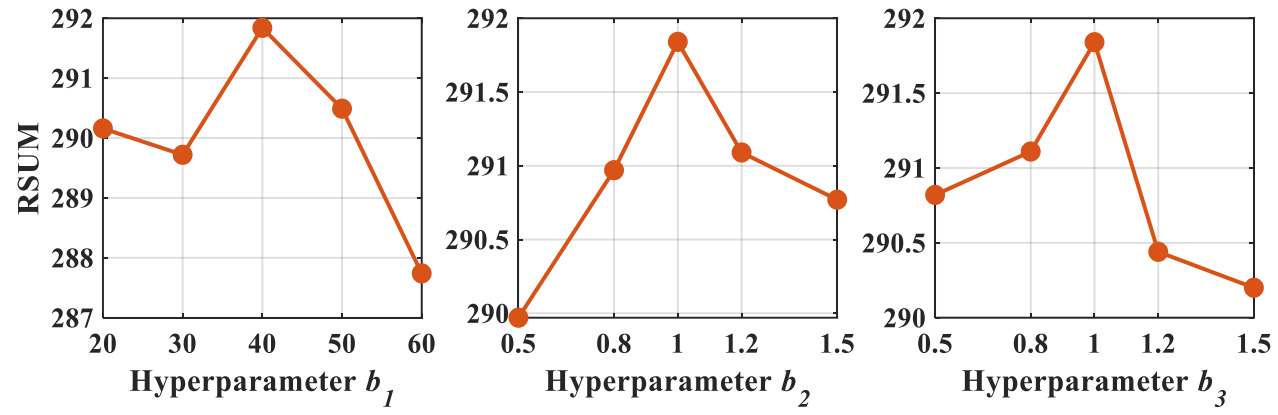

Fig. 10. Sensitivity analysis of hyperparameters on the RSITMD dataset.

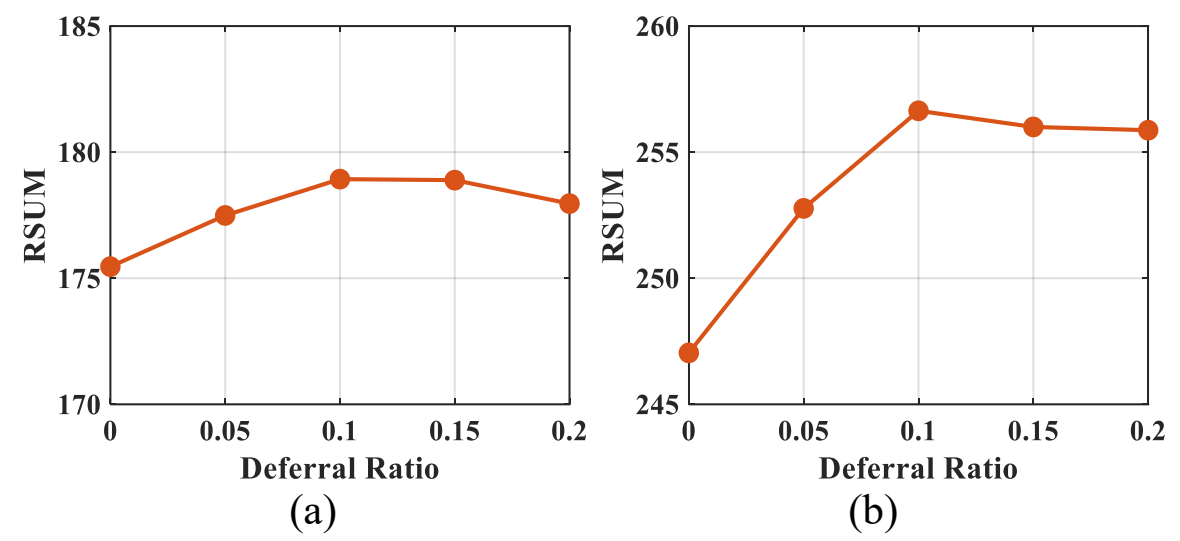

Fig. 11. Effect of the deferral ratios on the overall retrieval performance (RSUM). (a) Results on the Noisy RSICD-Test dataset. (b) Results on the Noisy RSITMD-Test dataset.

TABLE V

COMPARISON OF RS-AWARE TEST-TIME AUGMENTATION AGAINST GENERIC-TTA BASELINES ON NOISY RSICD-TEST DATASET.

| | Noisy RSICD-Test | | | | | | |
|---|---|---|---|---|---|---|---|
| | Image to Text | | | Text to Image | | | RSUM |
| | R@1 | R@5 | R@10 | R@1 | R@5 | R@10 | |
| **ELC0** | 14.82 | **31.02** | 43.66 | 10.44 | 31.16 | 44.36 | 175.46 |
| **ELC** | **15.28** | 30.56 | **45.38** | **10.83** | 31.24 | 45.64 | **178.93** |
| **ELC_nTTA** | 13.08 | 30.19 | 43.82 | 10.27 | **31.51** | **45.73** | 174.60 |

TABLE VI

COMPARISON OF RS-AWARE TEST-TIME AUGMENTATION AGAINST GENERIC-TTA BASELINES ON NOISY RSITMD-TEST DATASET.

| | Noisy RSITMD-Test | | | | | | |
|---|---|---|---|---|---|---|---|
| | Image to Text | | | Text to Image | | | RSUM |
| | R@1 | R@5 | R@10 | R@1 | R@5 | R@10 | |
| **ELC0** | 21.86 | 42.49 | 56.42 | 18.06 | **48.67** | 63.90 | 251.40 |
| **ELC** | **23.01** | **44.03** | **57.30** | **19.03** | 47.60 | **65.67** | **256.64** |
| **ELC_nTTA** | 22.34 | 43.14 | 56.20 | **19.03** | 47.55 | 65.10 | 253.36 |

RSITMD-Test datasets. The deferral ratio denotes the proportion of test queries with the highest estimated uncertainty that are selected for RS-TTA. Specifically, the ratio is varied in $\{0, 0.05, 0.1, 0.15, 0.2\}$. As illustrated in Fig. 11, the use of RS-TTA leads to consistent performance gains, i.e., augmenting high-uncertainty samples effectively mitigates semantic ambiguity caused by noisy inputs. As the ratio increases from 0 to 0.1, the RSUM consistently improves on both datasets and reaches its highest value at 0.1. However, when the ratio exceeds 0.1, the performance begins to degrade slightly, demonstrating that applying RS-TTA to an excessive number of less uncertain samples becomes counterproductive.

### *G. Effectiveness of RS-Aware Test-Time Augmentation*

The RS-TTA in ELC is invoked only for queries whose estimated uncertainty exceeds the deferral threshold. A natural question is whether the augmentation operators used in this deferral stage need to be RS-specific, or whether off-the-shelf augmentation operators from the natural-image literature would suffice. To answer this, we construct ELC_nTTA, a variant that shares the same trained checkpoint and the same TTA ratio as ELC but replaces the RS-aware augmentation set with generic operators, such as color jitter, random crop, horizontal flip, and EDA synonym replacement—commonly used in natural-image TTA.

Results on the Noisy RSICD-Test and Noisy RSITMD-Test datasets are reported in Tables V and VI. Across both datasets, ELC consistently outperforms ELC_nTTA, with RSUM gaps of +4.33 on Noisy RSICD-Test and +3.28 on Noisy RSITMD-Test. In addition, ELC_nTTA does not consistently improve over ELC0, which suggests that generic TTA augmentations may be unsuitable for CMRSITR because their transformations may disturb the radiometric appearance of RS images, modify spatial layouts, or replace RS-domain terms with semantically mismatched words, thereby weakening image-text correspondence. These results show that augmentation operators designed for RS imagery are better suited to CMRSITR than generic operators directly borrowed from the natural-image setting.

## V. CONCLUSION

In this paper, we proposed ELC to address the limitation that existing CMRSITR methods perform retrieval with full certainty for each test query and cannot exploit the per-query uncertainty induced by RS-specific test-time degradations. During training, EDL models inter-modal correspondences and estimates per-query uncertainty. UCL is introduced to encourage the uncertainty estimated by EDL to be associated with retrieval correctness, thereby enhancing uncertainty awareness. Moreover, RL preserves intra-modal similarity relationships from pretrained mentor encoders to support EDL-based uncertainty learning. In the test phase, ELC performs uncertainty-gated selective retrieval, where low-uncertainty queries are directly returned and high-uncertainty queries are refined by RS-TTA. Extensive experiments verified that the proposed method achieves competitive retrieval performance compared with state-of-the-art baseline methods and exhibits greater robustness to noise in test-time data. Although the uncertainty threshold is determined by a fixed deferral ratio in this study, adaptive thresholding may be further explored to make uncertainty-gated selective retrieval more flexible across deployment conditions in the future.